\documentclass{aa}  
\DeclareUnicodeCharacter{0301}{\'e}
\usepackage{graphicx}
\usepackage{txfonts}
\usepackage{multirow}
\usepackage{hyperref}
\usepackage{comment}

\usepackage{color}
\newcommand{\orcid}[1]{\href{https://orcid.org/#1}{\includegraphics[width=10pt]{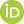}}}

\begin{document}

    \titlerunning{Analytical modeling of polarization signals arising from confined CSM in SNe II}
    \authorrunning{T. Nagao et al.}

   \title{Analytical modeling of polarization signals arising from confined circumstellar material in Type II supernovae
   }


   \author{T. Nagao\inst{1,2,3 \orcid{0000-0002-3933-7861}}\fnmsep\thanks{takashi.nagao@utu.fi}
          \and
          K. Maeda \inst{4\orcid{0000-0003-2611-7269}}
          \and
          T. Matsumoto \inst{5\orcid{0000-0002-9350-6793}}
          }

    \institute{
            Department of Physics and Astronomy, University of Turku, FI-20014 Turku, Finland
            \and
            Aalto University Mets\"ahovi Radio Observatory, Mets\"ahovintie 114, 02540 Kylm\"al\"a, Finland
            \and
            Aalto University Department of Electronics and Nanoengineering, P.O. BOX 15500, FI-00076 AALTO, Finland
            \and
            Department of Astronomy, Kyoto University, Kitashirakawa-Oiwake-cho, Sakyo-ku, Kyoto 606-8502, Japan
            \and
            Department of Astronomy, School of Science, The University of Tokyo, 7-3-1 Hongo, Bunkyo-ku, Tokyo 113-0033, Japan
             }

   \date{Received September 15, 1996; accepted March 16, 1997}

 
  \abstract
   {Recent observations of Type II supernovae (SNe) have brought a challenge in our understanding on the final evolutionary stage of massive stars. The early-time spectra and light curves of Type II SNe suggest that a majority of them have dense circumstellar material (CSM) in their vicinity, the so-called confined CSM. However, the mechanism of these extensive mass loss has not yet been understood.
   }
   {For addressing this problem, we aim to study the spatial distribution of the confined CSM, which has important information on the mechanism.
   }
   {We analytically calculate the polarization signals created by electron scatterings within disk-like confined CSM, and apply the results to the case of SN~2023ixf.}
   {The calculated polarization angle remains fixed at the angle aligned with the CSM disk axis, and is insensitive to the disk parameters. The calculated polarization degree evolves over a timescale of $\lesssim 10$ days, depending on the disk parameters: it stays constant or increases slightly while the unshocked CSM is optically thick, peaks as it becomes optically thin, and drops to zero when the shock reaches the disk’s outer edge. We also find that the time evolution of the polarization in Type II SNe with confined CSM can be used for estimating the CSM parameters. In particular, the maximum degree and the rise time are strongly connected to the values of the viewing angle and the opening angle of the CSM disk, while the duration and the decline time are sensitive to the values of the mass and extension of the CSM disk. We demonstrate that the time evolution of the polarization of SN 2023ixf can be explained with a disk-like CSM with the following parameters: the viewing angle of $\theta_{\rm{obs}} \gtrsim 40$ degrees, the half-opening angle of the disk of $\theta_{\rm{disk}} \sim 50-60$ degrees, the CSM mass of $M_{\rm{csm}}\sim 2\times10^{-3}$ M$_{\odot}$, and the outer edge of the CSM disk of $r_{\rm{out}} \sim 3\times 10^{14}$ cm.}
   {This information of the CSM is a strong constraint on the mechanism to create the confined CSM. Moreover, the observed alignment between the explosion asymmetry and the CSM disk of SN~2023ixf may point to a shared origin of these structures, possibly associated with the progenitor star itself rather than with companion interaction. Further early-time polarimetry of Type II SNe will be crucial for clarifying the underlying mechanism by probing the diversity of confined CSM. In addition, since this method can be applied to calculations of polarization originating from arbitrary shape of aspherical scattering-dominated photospheres, we can study the geometries of a wide range of objects with scattering-dominated photospheres using this method.}

   \keywords{giant planet formation --
                $\kappa$-mechanism --
                stability of gas spheres
               }

   \maketitle
%

\section{Introduction}

Type II supernovae (SNe) are explosions of red supergiant stars, whose initial zero-age-main-sequence masses are between $\sim8$ and $\sim18$ M$_{\odot}$ \citep[e.g.,][]{Smartt2015,VanDyk2025}. Recent observations of Type II SNe have brought a challenge in our understanding on the final evolutionary stages of massive stars. Their early-phase spectra and light curves suggest that a majority of Type II SNe have dense circumstellar material (CSM) in their vicinity ($\lesssim 10^{15}$ cm; so-called confined CSM), corresponding to mass-loss rates of $\sim 10^{-4} - 1$ M$_{\odot}$ yr$^{-1}$ \citep[e.g.,][]{Khazov2016, Yaron2017, Forster2018, Boian2020, Bruch2021}. However, these extensive mass eruptions cannot be quantitatively explained by the currently proposed mechanisms, e.g., mass eruptions due to some stellar instabilities \citep[e.g.,][]{Humphreys1994, Langer1999, Yoon2010, Arnett2011, Quataert2012, Shiode2014, Smith2014, Woosley2015, Quataert2016, Fuller2017, Sengupta2025} or binary interaction \citep[e.g.,][]{Chevalier2012, Soker2013}.

Deriving the spatial distribution of the confined CSM is important for clarifying the mass-loss mechanism to create it. The early-time polarimetry of Type II SN~2023ixf revealed that its confined CSM is aspherical, showing the continuum polarization of $\sim 1$ \%, within the first several days \citep[][]{Vasylyev2023}.
In this paper, we analytically calculate the polarization signals that originate from the confined CSM in Type II SNe, in order to determine the precise distribution of the CSM by apllying the calculations to polarimetric observations. In Section~\ref{sec:calculations}, we describe our calculation methods for the polarization signals created by the electron scattering processes within the CSM. In Section~\ref{sec:results}, we provide the results and discuss the dependence of the polarization on the parameters of the confined CSM. In Section~\ref{sec:limitation}, we discuss the limitations of our calculations. We conclude the paper with discussions in Section~\ref{sec:conclusion}.

\section{Calculations} \label{sec:calculations}

\begin{figure}
   \centering
   \includegraphics[width=\hsize]{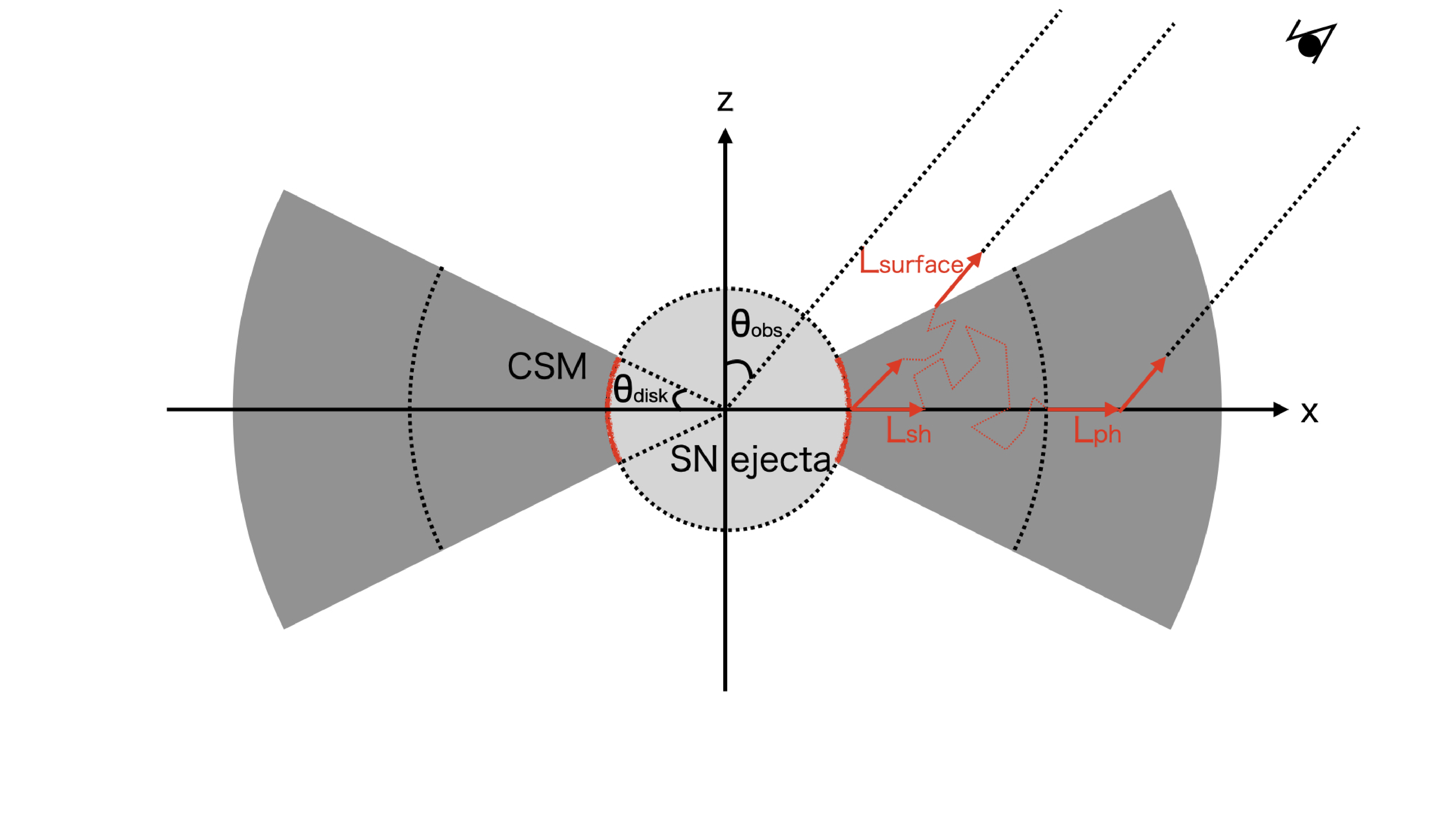}
      \caption{Schematic picture of our calculations.
              }
         \label{fig:fig1}
\end{figure}

In this section, we describe our methodology to calculate the polarization signals that originate from the confined CSM. Figure~\ref{fig:fig1} shows a schematic picture of our calculations. For the CSM, we consider a disk-shaped structure whose inner and outer edges are at the surface of the progenitor star ($R_{\rm{p}}$) and at a radius of $r_{\rm{out}}$, respectively, with a half-opening angle of $\theta_{\rm{disk}}$. The observer is located on the $x$–$z$ plane at an angle $\theta_{\rm{obs}}$ from the $z$-axis.
We assume that the radiation is created only by the interaction between the SN ejecta and the confined CSM, ignoring the radiation from the SN ejecta. This can be justified only for early phases of Type II SNe \citep[for the first few weeks; e.g.,][]{Ertini2025}. We also assume, for simplicity, that the confined CSM is fully ionized and that electron scattering within the CSM is the only radiation process.

First, we calculate the radiation generated by the CSM interaction using the analytical formula from \citet[][]{Moriya2013}, based on the assumed SN ejecta and CSM properties. Then, we divide the radiation into two components: the radiation that eventually escapes from the photosphere in the radial direction of the CSM disk and the one that escapes from the photosphere at the surface of the CSM disk (see Figure~\ref{fig:fig1}), treating the polarization of these components separately. Once the unshocked CSM becomes optically thin, we treat all the radiation as escaping from the photosphere in the interaction region.

\subsection{SN ejecta and CSM properties}

As the SN ejecta properties, we assume that the SN ejecta expand homologously and the density profile of the SN ejecta follows the outer part of the double power-law distribution proposed by numerical calculations of SN explosions (\citet[][]{Matzner1999}, see also \citet[][]{Moriya2013}):
\begin{eqnarray}
\label{eq:rho_ej}
\rho_{\rm{ej}} (v_{\rm{ej}},t) = 
      \frac{1}{4 \pi (n-\delta)} \frac{[2(5-\delta)(n-5) E_{\rm{ej}}]^{\frac{n-3}{2}}}{[(3-\delta)(n-3) M_{\rm{ej}}]^{\frac{n-5}{2}}} t^{-3} v_{\rm{ej}}^{-n}.
\end{eqnarray}
Here, $v_{\rm{ej}} (r,t)$ ($= r/t$) is the velocity of the SN ejecta at radius, $r$, and time, $t$. We assume that $n=12$ and $\delta =1$, which are typical values for the outer ejecta of a red supergiant progenitor \citep[e.g.,][]{Matzner1999}. Throughout the paper, we also assume that the ejecta mass ($M_{\rm{ej}}$) and energy ($E_{\rm{ej}}$) to be 10 M$_{\odot}$ and $10^{51}$ erg, respectively, which are typical values for Type IIP SNe \citep[e.g.,][]{Martinez2022}.

We assume that the density of the CSM depends solely on the distance from the center of the SN ejecta, and that its radial distribution within the disk follows a power law, $\rho_{\rm{disk}} (r) = D r^{-s}$. 
Instead of the parameter $D$, we use the total mass of the disk CSM ($M_{\rm{csm}}$) to characterize the CSM density. The total mass of the CSM is related to $D$ as follows:
\begin{equation}
\label{eq:M_disk}
M_{\rm{disk}} = \int^{r_{\rm{out}}}_{R_{\rm{p}}} 4 \pi \omega_{\rm{disk}} r^{2} \rho_{\rm{disk}}(r) \mathrm{d} r = \frac{4 \pi \omega_{\rm{disk}} D}{3-s} \left( r_{\rm{out}}^{3-s}-R_{\rm{p}}^{3-s} \right),
\end{equation}
where $\omega_{\rm{disk}}=\Omega_{\rm{disk}}/4\pi=\sin \theta_{\rm{disk}}$, and $\Omega_{\rm{disk}}$ is the solid angle subtended by the CSM disk covering the SN ejecta.
Thus, we get the following equation:
\begin{eqnarray}
\label{eq:D_M_csm}
D = \frac{(3-s)M_{\rm{disk}}}{4 \pi \omega_{\rm{disk}} (r_{\rm{out}}^{3-s}-R_{\rm{p}}^{3-s})}.
\end{eqnarray}

In the case of $s=2$, we can define the corresponding mass-loss rate to create the disk CSM as follows:
\begin{eqnarray}
\label{eq:D_M_csm}
\dot{M}_{\rm{disk}} &=& \frac{M_{\rm{disk}} v_{\rm{csm}}}{r_{\rm{out}}- R_{\rm{p}}}, \nonumber\\
&\sim& 1 \times 10^{-1} \left( \frac{M_{\rm{disk}}}{10^{-1} [\rm{M}_{\odot}]} \right) \left( \frac{v_{\rm{csm}}}{100 [\rm{km}\;\rm{s}^{-1}] } \right) \left( \frac{r_{\rm{out}}}{3\times 10^{14} [\rm{cm}] } \right)^{-1} \nonumber\\
&& \;\;\;\;\;\;\;\;\;\;\;\;\;\;\;\;\;\;\;\;\;\;\;\;\;\;\;\;\;\;\;\;\;\;\;\;\;\;\;\;\;\;\;\;\;\;\;\;\;\;\;\;\;\;\;\;\;\;\; [\rm{M}_{\odot}\;\rm{yr}^{-1}].
\end{eqnarray}
Here, we assume that $r_{\rm{out}} >> R_{\rm{p}}$. We also assume that the velocity of the CSM ($v_{\rm{csm}}$) is 100 km s$^{-1}$ throughout the paper.

\subsection{CSM interaction}

We calculate the bolometric luminosity created by the interaction between the SN ejecta and CSM disk ($L_{\rm{sh}}$), using the analytical model proposed by \citet[][]{Moriya2013}. The time evolution of the location ($r_{\rm{sh}}(t)$) and velocity ($v_{\rm{sh}}(t)$) of the shocked shell can be calculated by solving the equation of motion of the shocked shell, assuming its width is negligible compared to its radius:
\begin{eqnarray}
\label{eq:eom}
    M_{\rm{sh}} (t) \frac{dv_{\rm{sh}}(t)}{dt} &=& 4 \pi \omega_{\rm{disk}} r_{\rm{sh}}^{2}(t) \Bigl[ \rho_{\rm{ej}} (r_{\rm{sh}}(t),t) \left( v_{\rm{ej}}(r_{\rm{sh}}(t),t)-v_{\rm{sh}}(t) \right)^{2} \nonumber\\
    && - \rho_{\rm{disk}} (r_{\rm{sh}}(t)) \left( v_{\rm{sh}}(t)-v_{\rm{csm}} \right)^{2} \Bigr].
\end{eqnarray}
Here, $v_{\rm{sh}}(t)$ is the velocity of the shocked shell at time of $t$. $M_{\rm{sh}}(t)$ is the total mass of the shocked shell at time of $t$, shown as follows:
\begin{equation}
    M_{\rm{sh}} (t) = \int_{R_{\rm{p}}}^{r_{\rm{sh}}(t)} 4 \pi \omega_{\rm{disk}} r^{2} \rho_{\rm{disk}} (r) \mathrm{d}r + \int_{r_{\rm{sh}}(t)}^{r_{\rm{ej,max}}(t)} 4 \pi r^{2} \omega_{\rm{disk}} \rho_{\rm{ej}} (r,t) \mathrm{d}r
\end{equation}
where $r_{\rm{ej,max}}(t) = v_{\rm{ej,max}} t$, and $v_{\rm{ej,max}}$ is the original velocity of the outermost layer of the SN ejecta before the CSM interaction. Here, we assume $r_{\rm{sh}}(t) >> R_{\rm{p}}$ and $r_{\rm{ej,max}}(t) >> r_{\rm{sh}}(t)$.

Using Eq.~\eqref{eq:eom}, we can analytically calculate the time-evolution of the shocked shell and its velocity as follows (see \citet{Moriya2013}):
\begin{equation}
\label{eq:r_sh}
r_{\rm{sh}} (t) = \left[ \frac{(3-s)(4-s)}{4\pi D(n-4)(n-3)(n-\delta)} \frac{[2(5-\delta)(n-5)E_{\rm{ej}}]^{(n-3)/2}}{[(3-\delta)(n-3)M_{\rm{ej}}]^{(n-5)/2}} \right]^{\frac{1}{n-s}} t^{\frac{n-3}{n-s}},
\end{equation}
and
\begin{eqnarray}
\label{eq:v_sh}
v_{\rm{sh}} (t) &=& \frac{n-3}{n-s} \left[ \frac{(3-s)(4-s)}{4\pi D(n-4)(n-3)(n-\delta)} \frac{[2(5-\delta)(n-5)E_{\rm{ej}}]^{(n-3)/2}}{[(3-\delta)(n-3)M_{\rm{ej}}]^{(n-5)/2}} \right]^{\frac{1}{n-s}} \nonumber\\
&& \;\;\;\;\;\;\;\;\;\;\;\;\;\;\;\;\;\;\;\;\;\;\;\;\;\;\;\;\;\;\;\;\;\;\;\;\;\;\;\;\;\;\;\;\;\;\;\;\;\;\;\;\;\;\;\;\;\;\;\;\;\;\;\;\;\;\;\;\;\;\; t^{-\frac{3-s}{n-s}}.
\end{eqnarray}

We assume the bolometric luminosity from the shocked shell to be a fraction of the released energy by the shock, as follows:
\begin{equation}
    L_{\mathrm{sh}}(t) = \epsilon \frac{\mathrm{d}E_{\rm{kin}}(t)}{\mathrm{d}t} = 2 \pi \omega_{\rm{disk}} \epsilon \rho_{\mathrm{disk}} (r_{\mathrm{sh}}(t)) r_{\mathrm{sh}}^{2}(t) v_{\mathrm{sh}}^{3}(t),
\end{equation}
where $\epsilon$ is the conversion efficiency from kinetic energy to radiation, and
\begin{equation}
    \mathrm{d}E_{\mathrm{kin}}(t) = 4 \pi \omega_{\mathrm{disk}} r_{\mathrm{sh}}^{2}(t) \left( \frac{1}{2} \rho_{\mathrm{disk}} (r_{\mathrm{sh}}(t)) v_{\mathrm{sh}}^{2}(t) \right) \mathrm{d}r.
\end{equation}
We note that the choice of the value of $\epsilon$ does not affect the calculated polarization degrees.
%
Then, the luminosity from the photosphere and surface of the optically-thick region of the disk is simply assumed to be determined by the relative optical depths of the CSM disk along the radial and perpendicular directions ($\tau_{\rm{csm,r}}$ and $\tau_{\rm{csm,h}}$, respectively) without taking the diffusion time:
\begin{eqnarray}
L_{\rm{ph}} (t) = \left\{
    \begin{array}{l}
      L_{\rm{sh}} (t) \times \frac{\tau_{\rm{csm,h}}(r_{\rm{sh}})}{\tau_{\rm{csm,r}}(r_{\rm{sh}})+\tau_{\rm{csm,h}}(r_{\rm{sh}})} \;\;\;\; (\tau_{\rm{csm,r}} (r_{\rm{sh}}) \geq 1)\\
      L_{\rm{sh}} (t) \;\;\;\;\;\;\;\;\;\;\;\;\;\;\;\;\;\;\;\;\;\;\;\;\;\;\;\;\;\;\;\;\;\;\; (\tau_{\rm{csm,r}} (r_{\rm{sh}}) < 1)
    \end{array}
  , \right.
\end{eqnarray}
and
\begin{eqnarray}
L_{\rm{surface}} (t) = \left\{
    \begin{array}{l}
      L_{\rm{sh}} (t) \times \frac{\tau_{\rm{csm,r}}(r_{\rm{sh}})}{\tau_{\rm{csm,r}}(r_{\rm{sh}})+\tau_{\rm{csm,h}}(r_{\rm{sh}})} \;\;\;\; (\tau_{\rm{csm,r}}(r_{\rm{sh}}) \geq 1)\\
      0 \;\;\;\;\;\;\;\;\;\;\;\;\;\;\;\;\;\;\;\;\;\;\;\;\;\;\;\;\;\;\;\;\;\;\;\;\;\;\;\;\;\; (\tau_{\rm{csm,r}} (r_{\rm{sh}}) < 1)
    \end{array}
  . \right.
\end{eqnarray}
Here, we simply define the reference optical depths ($\tau_{\rm{csm,r}}$ and $\tau_{\rm{csm,h}}$) as the optical depths along the radial direction and along the inner radius of the CSM disk, respectively, as follows:
\begin{eqnarray}
\label{eq:tau_csm}
    \tau_{\rm{csm},r} (r) &=& \int^{r_{\rm{out}}}_{r} \kappa_{\rm{es}} D r^{-s} \mathrm{d}r, \nonumber\\
    &=&\frac{\kappa_{\rm{es}} D}{s-1} \left( r^{-(s-1)} - r_{\rm{out}}^{-(s-1)} \right).\\
    \tau_{\rm{csm},h} (r) &=& \kappa_{\rm{es}} \rho_{\rm{csm}} (r) r \theta_{\rm{disk}}, \nonumber\\
    &=& \kappa_{\rm{es}} D r^{-(s-1)} \theta_{\rm{disk}},
\end{eqnarray}
and the location of the photosphere is determined by $\tau_{\rm{csm},r} (r_{\rm{ph}})=1$ when $\tau_{\rm{csm,r}}(r_{\rm{sh}}) \geq 1$; otherwise, $r_{\rm{ph}}=r_{\rm{sh}}$.

In cases with $\theta_{\rm{obs}} \leqq \pi/2 - \theta_{\rm{disk}}$, the observer receive the radiation from the photosphere ($L_{\rm{ph}}$) as well as the one from the surface of the CSM disk ($L_{\rm{surface}}$; see Figure~\ref{fig:fig1}). In cases with $\theta_{\rm{obs}} > \pi/2 - \theta_{\rm{disk}}$, the radiation from the surface does not reach the observer, and thus we only consider the radiation from the photosphere.

\subsection{Polarization from the confined CSM}

The radiation reaching the observer can be divided into two components: (1) photons coming directly from the photosphere in the radial direction of the CSM disk, as well as photons initially emitted from the photosphere in directions out of the line of sight but subsequently scattered into the observer's direction, and (2) photons originating from the photosphere at the surface of the CSM disk (see Figure~\ref{fig:fig1}). We consider the polarization of these components separately in the following subsections. The Stokes parameters of the first component are denoted as $<I_{\rm{ph}}>$ and $<Q_{\rm{ph}}>$, and the second component as $<I_{\rm{surface}}>$ and $<Q_{\rm{surface}}>$. These Stokes parameters are defined in the Cartesian x-y-z coordinate, where the observer is on the x-z surface (see Figure~\ref{fig:fig2}).
The Stokes parameters and polarization degree and angle of the total radiation reaching to the observer are calculated as follows:

\begin{eqnarray}
    q_{\rm{obs}} &=& \frac{<Q_{\rm{ph}}>+<Q_{\rm{surface}}>}{<I_{\rm{ph}}>+<I_{\rm{surface}}>},\\
    u_{\rm{obs}} &=& \frac{<U_{\rm{ph}}>+<U_{\rm{surface}}>}{<I_{\rm{ph}}>+<I_{\rm{surface}}>},
\end{eqnarray}
and
\begin{eqnarray}
P &=& \sqrt{q_{\rm{obs}}^{2}+u_{\rm{obs}}^{2}}, \label{eq:pol_deg}\\
\theta_{P} &=& \frac{1}{2} \tan^{-1} \left( \frac{u_{\rm{obs}}}{q_{\rm{obs}}} \right). \label{eq:pol_angle}
\end{eqnarray}

\subsubsection{Radiation from the photosphere in the radial direction of the CSM disk ($\tau_{\rm{csm,r}} (r_{\rm{sh}}) > 1$)}

\begin{figure}
   \centering
   \includegraphics[width=\hsize]{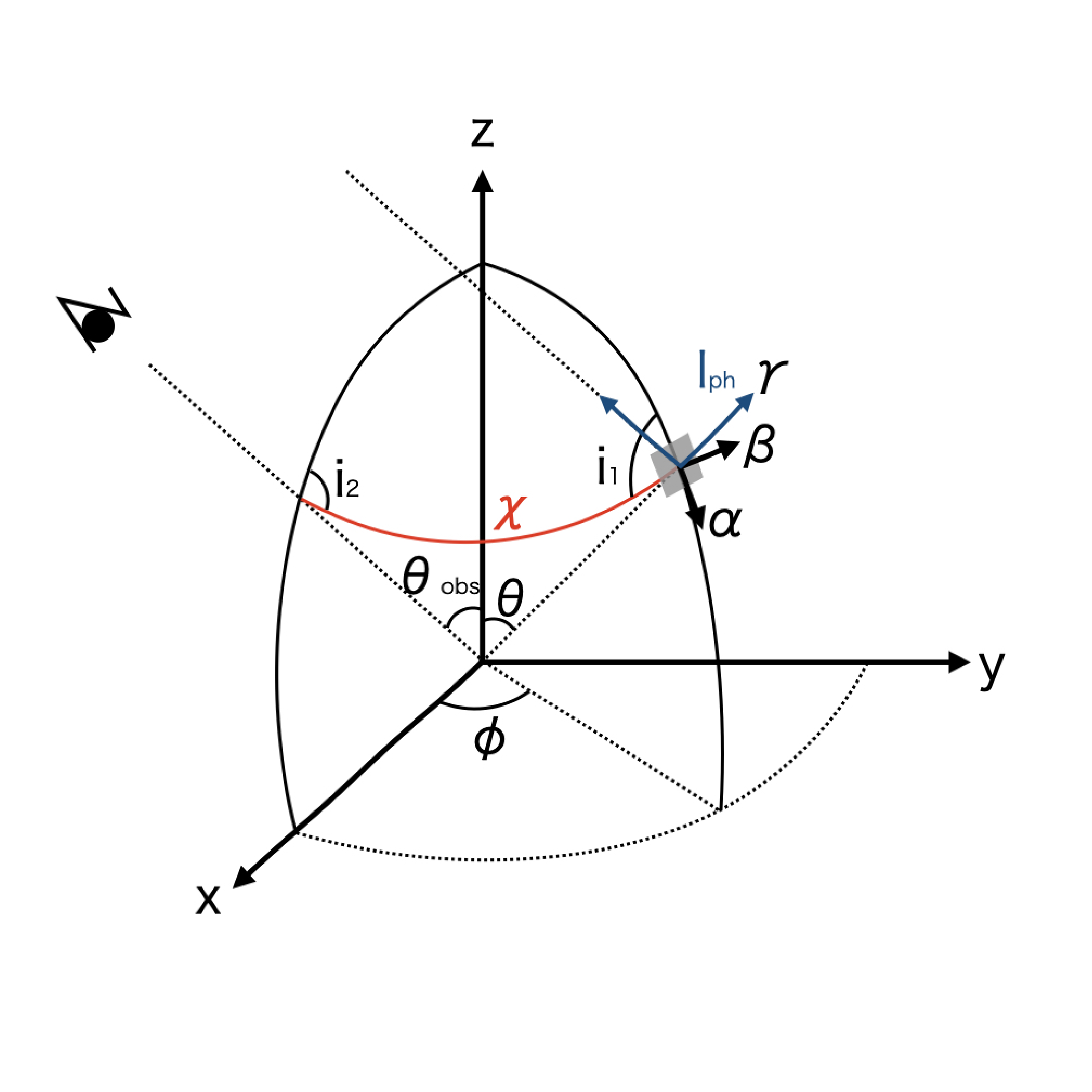}
      \caption{Geometry for scattering. Photons emerging from a differential surface area (gray region; $\theta$ and $\phi$ in the observer's frame) and scattered into the observer's direction at an angle $\chi$ are observed on the x-z surface with an angle of $\theta_{\rm{obs}}$ from the z axis.
              }
         \label{fig:fig2}
\end{figure}

We calculate the Stokes parameters of the radiation from the photosphere in the radial direction of the CSM disk by integrating the contributions from each differential surface area across the photosphere.
In this subsection, we consider cases with $\tau_{\rm{csm,r}}(r_{\rm{sh}}) > 1$, where there is a photosphere in the unshocked CSM. The flux and intensity of the radiation from the photosphere with the luminosity $L_{\rm{ph}}$ are written as follows:
\begin{eqnarray}
    f_{\rm{ph}} &=& \frac{L_{\rm{ph}}}{\int^{2\pi}_{0} d\phi \int^{\pi/2+\theta_{\rm{disk}}}_{\pi/2-\theta_{\rm{disk}}} d\theta \left( r_{\rm{ph}}^{2} \sin \theta \right)}, \nonumber\\
    &=& \frac{L_{\rm{ph}}}{4 \pi r_{\rm{ph}}^{2} \sin \theta_{\rm{disk}}},
\end{eqnarray}
\begin{eqnarray}
    I_{\rm{ph}} &=& \frac{f_{\rm{ph}}}{\int^{2\pi}_{0} d\phi \int^{\pi/2}_{0} d\theta \left( \sin \theta \right)}, \nonumber\\
    &=& \frac{L_{\rm{ph}}}{8 \pi^{2} r_{\rm{ph}}^{2} \sin \theta_{\rm{disk}}}.
\end{eqnarray}

We assume that all photons emerging from the photosphere in the radial direction of the CSM disk are unpolarized and are subsequently scattered once above the photosphere into random directions, including cases that do not change the directions of photons before and after the scattering. Photons scattered into the observer's direction are detected.
First, we consider such photons from the differential surface area at ($\theta$, $\phi$) in the Cartesian $x$-$y$-$z$ coordinate (see Figure~\ref{fig:fig2}). Here, the observer is on the $x$-$z$ surface with the angle of $\theta_{\rm{obs}}$ from the $z$ axis. The spherical $\alpha$-$\beta$-$\gamma$ coordinate is defined toward the direction normal to the differential surface area at ($\theta$,$\phi$) on the photosphere, as in Figure~\ref{fig:fig2}. 
%

The Stokes parameters of such photons from a differential surface area at ($\theta$, $\phi$) in the spherical $\alpha$-$\beta$-$\gamma$ coordinate ($I_{\rm{total}}$, $Q_{\rm{total}}$, and $U_{\rm{total}}$) can be calculated as shown in Appendix~\ref{sec:app1}:
\begin{equation} \label{eq:final_Stokes_para}
    \begin{cases}
    I_{\rm{total}} (\chi) = I_{\rm{ph}} \left( \frac{9}{8} - \frac{3}{16} \sin^{2} \chi \right),\\
    Q_{\rm{total}} (\chi) = -\frac{3}{16} I_{\rm{ph}} \sin^{2} \chi,\\
    U_{\rm{total}} (\chi) = 0,
    \end{cases}
\end{equation}
where $\chi$ is the angle between the unit vector ${\bf n}$ normal to the differential surface area at ($\theta$,$\phi$) on the photosphere and the the unit vector for the observer's direction, {\bf $n_{\rm{obs}}$} (see also Figure~\ref{fig:fig2}). Using the law of cosines in spherical trigonometry, this angle is expressed as
\begin{equation}
    \cos \chi = {\bf n} \cdot {\bf n_{\rm{obs}}} = \sin \theta_{\rm{obs}} \sin \theta \cos \phi + \cos \theta_{\rm{obs}} \cos \theta.
\end{equation}

The Stokes parameters of these photons in the observer's frame ($I_{\rm{obs-frame}}$, $Q_{\rm{obs-frame}}$, and $U_{\rm{obs-frame}}$)
are derived by rotating the Stokes parameters defined in the spherical $\alpha-\beta-\gamma$ coordinate ($I_{\rm{total}}$, $Q_{\rm{total}}$, and $U_{\rm{total}}$) with angle $\chi$ as follows:
\begin{eqnarray}
    \left(
    \begin{array}{c}
    I_{\rm{obs-frame}}\\
    Q_{\rm{obs-frame}} \\
    U_{\rm{obs-frame}} 
    \end{array}
\right) &=& \mathbb{L} (\pi-i_{2})
\left(
    \begin{array}{c}
    I_{\rm{total}}\\
    Q_{\rm{total}} \\
    U_{\rm{total}} 
    \end{array}
\right) \nonumber\\
&=& \left(
    \begin{array}{c}
    I_{\rm{total}}\\
    Q_{\rm{total}} \cos 2 i_{2} - U_{\rm{total}} \sin 2 i_{2}\\
    Q_{\rm{total}} \sin 2 i_{2} + U_{\rm{total}} \cos 2 i_{2}
    \end{array}
\right).
\end{eqnarray}
Here, $\mathbb{L} (\psi)$ is the rotation matrix:
\begin{eqnarray}
    \mathbb{L} (\psi) = \left(
    \begin{array}{ccc}
    1 & 0 & 0 \\
    0 & \cos 2\psi & \sin 2\psi \\
    0 & -\sin 2\psi & \cos 2\psi
    \end{array}
\right).
\end{eqnarray}

\begin{figure*}
   \centering
   \includegraphics[width=0.49\hsize]{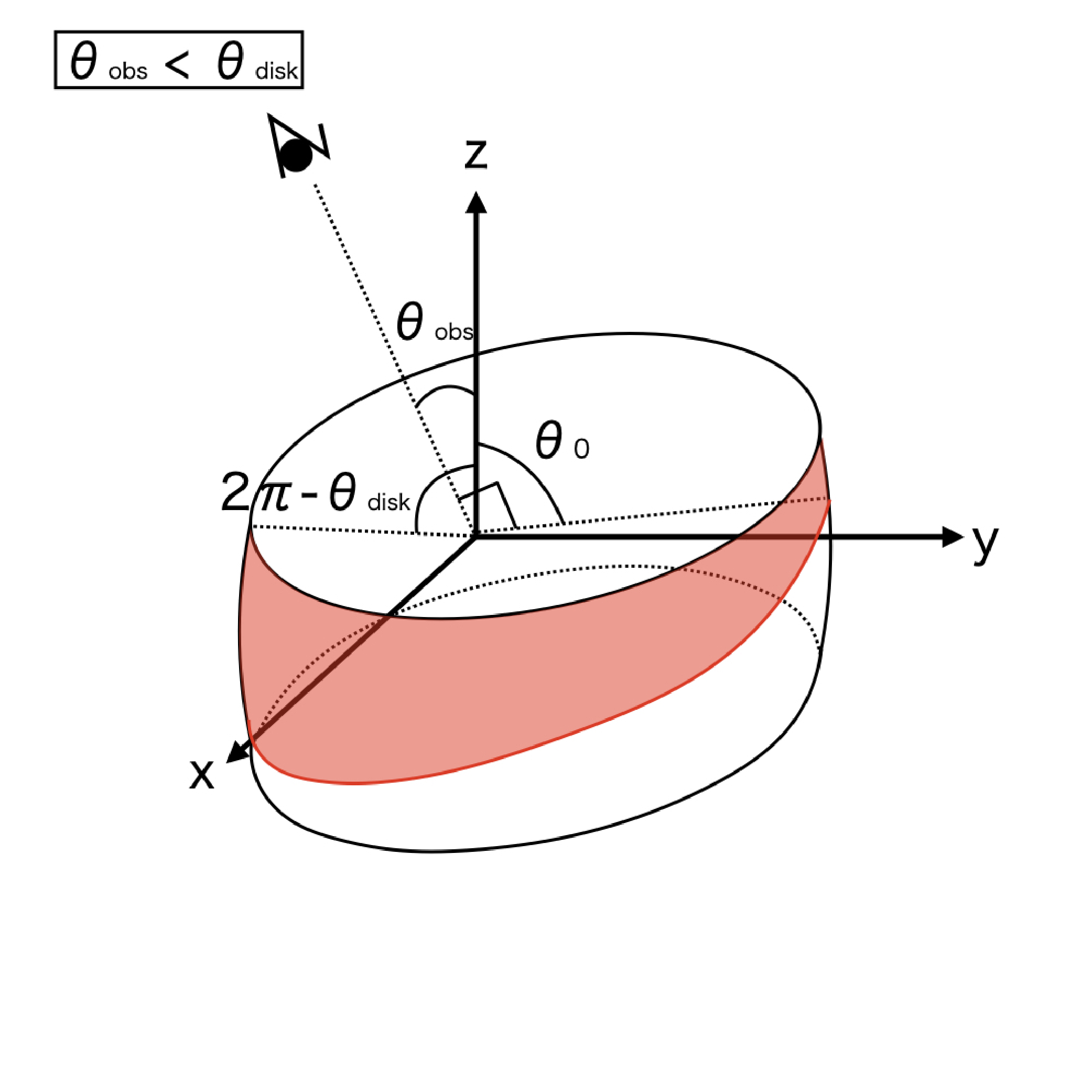}
   \includegraphics[width=0.49\hsize]{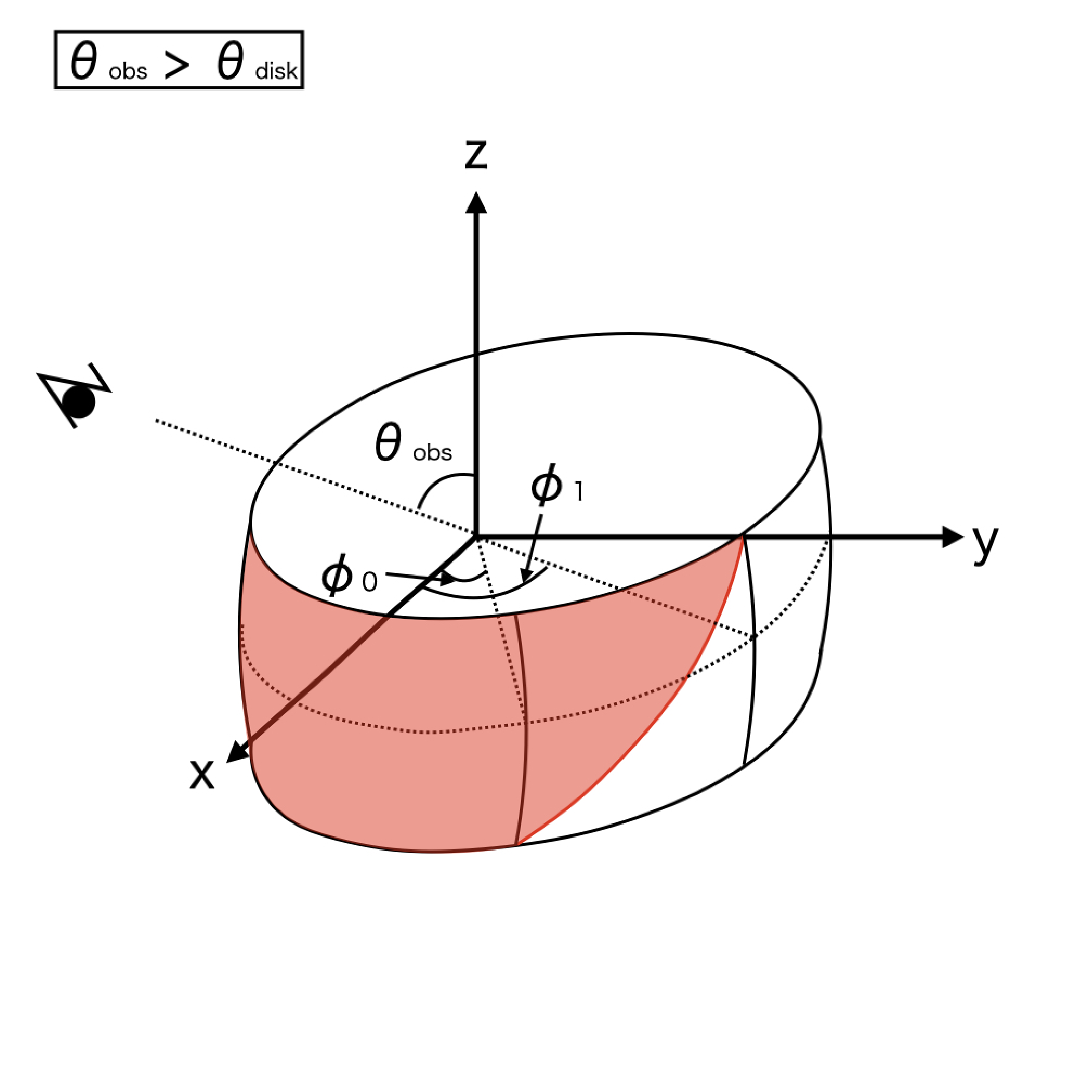}
      \caption{Schematic illustration for the integration of the Stokes parameters. The red hatches show the areas of the photosphere on the disk edge visible to the observer for cases with different combinations of $\theta_{\rm{obs}}$ and $\theta_{\rm{disk}}$.
              }
         \label{fig:fig3}
   \end{figure*}

Then, we derive the Stokes parameters of the radiation from the photosphere in the radial direction of the CSM disk ($<I_{\rm{ph}}>$, $<Q_{\rm{ph}}>$, and $<U_{\rm{ph}}>$, defined in the observer's frame, i.e., in the Cartesian $x$-$y$-$z$ coordinate) by integrating all contributions from each differential surface area across the photosphere visible to the observer.
The integration area is the photosphere above the surface perpendicular to the observer's direction, {\bf $n_{\rm{obs}}$}, passing through the origin of the $x$-$y$-$z$ coordinate system (see Figure~\ref{fig:fig3}). Here, the value of the polar angle ($\theta_0$) corresponding to the direction perpendicular to the observer’s line of sight, at a given azimuthal angle ($\phi$), is derived from the right-angle relationship as follows:
\begin{equation}
    \tan \theta_0 = - \frac{1}{cos \phi} \frac{1}{\tan \theta_{\rm{obs}}} = \frac{\tan (\theta_{\rm{obs}}+\pi/2)}{cos \phi}.
\end{equation}

In cases with $\theta_{\rm{obs}} < \theta_{\rm{disk}}$, we thus derive the Stokes parameters as follows:
\begin{eqnarray}
    <I_{\rm{ph}}> = \int^{2\pi}_{0} d\phi \int^{\theta_{0}}_{\pi/2-\theta_{\rm{disk}}} d\theta \left( I_{\rm{obs-frame}} \sin\theta \right), \nonumber\\
    <Q_{\rm{ph}}> = \int^{2\pi}_{0} d\phi \int^{\theta_{0}}_{\pi/2-\theta_{\rm{disk}}} d\theta \left( Q_{\rm{obs-frame}} \sin\theta \right), \nonumber\\
    <U_{\rm{ph}}> = \int^{2\pi}_{0} d\phi \int^{\theta_{0}}_{\pi/2-\theta_{\rm{disk}}} d\theta \left( U_{\rm{obs-frame}} \sin\theta \right), \nonumber
\end{eqnarray}

In the cases where $\theta_{\rm{obs}} > \theta_{\rm{disk}}$, we define the azimuthal angles of the crossing points between the photosphere and the bottom ($\theta=\pi/2+\theta_{\rm{disk}}$) and top ($\theta=\pi/2-\theta_{\rm{disk}}$) disk corners as $\phi_0$ and $\phi_1$, respectively (see Figure~\ref{fig:fig3}). We calculate the Stokes parameters as follows:
\begin{eqnarray}
    <I_{\rm{ph}}> &=& \int^{\phi_{0}}_{0} d\phi \int^{\pi/2+\theta_{\rm{disk}}}_{\pi/2-\theta_{\rm{disk}}} d\theta \left( I_{\rm{obs-frame}} \sin\theta \right) \nonumber\\
    &&+ \int^{\phi_{1}}_{\phi_{0}} d\phi \int^{\theta_{0}}_{\pi/2-\theta_{\rm{disk}}} d\theta \left( I_{\rm{obs-frame}} \sin\theta \right) \nonumber\\
    &&+ \int^{2\pi-\phi_{0}}_{2\pi-\phi_{1}} d\phi \int^{\theta_{0}}_{\pi/2-\theta_{\rm{disk}}} d\theta \left( I_{\rm{obs-frame}} \sin\theta \right) \nonumber\\
    &&+ \int^{2\pi}_{2\pi-\phi_{0}} d\phi \int^{\pi/2+\theta_{\rm{disk}}}_{\pi/2-\theta_{\rm{disk}}} d\theta \left( I_{\rm{obs-frame}} \sin\theta \right), \nonumber\\
    &=& 2\int^{\phi_{0}}_{0} d\phi \int^{\pi/2+\theta_{\rm{disk}}}_{\pi/2-\theta_{\rm{disk}}} d\theta \left( I_{\rm{obs-frame}} \sin\theta \right)\\
    &&+ 2\int^{\phi_{1}}_{\phi_{0}} d\phi \int^{\theta_{0}}_{\pi/2-\theta_{\rm{disk}}} d\theta \left( I_{\rm{obs-frame}} \sin\theta \right), \nonumber\\
    <Q_{\rm{ph}}> &=& 2\int^{\phi_{0}}_{0} d\phi \int^{\pi/2+\theta_{\rm{disk}}}_{\pi/2-\theta_{\rm{disk}}} d\theta \left( Q_{\rm{obs-frame}} \sin\theta \right)\\
    &&+ 2\int^{\phi_{1}}_{\phi_{0}} d\phi \int^{\theta_{0}}_{\pi/2-\theta_{\rm{disk}}} d\theta \left( Q_{\rm{obs-frame}} \sin\theta \right), \nonumber\\
    <U_{\rm{ph}}> &=& 2\int^{\phi_{0}}_{0} d\phi \int^{\pi/2+\theta_{\rm{disk}}}_{\pi/2-\theta_{\rm{disk}}} d\theta \left( U_{\rm{obs-frame}} \sin\theta \right)\\
    &&+ 2\int^{\phi_{1}}_{\phi_{0}} d\phi \int^{\theta_{0}}_{\pi/2-\theta_{\rm{disk}}} d\theta \left( U_{\rm{obs-frame}} \sin\theta \right). \nonumber\\
\end{eqnarray}
In both cases, we get the following formulas:
\begin{eqnarray}
    <I_{\rm{ph}}> &=& \frac{\pi I_{\rm{ph}}}{16} [ 3(11- \cos^2 \theta_{\rm{obs}}) \sin \theta_{\rm{disk}} \nonumber\\
    && - (1-3\cos^2 \theta_{\rm{obs}}) \sin^3 \theta_{\rm{disk}} ] \label{eq:I_ph}\\
    <Q_{\rm{ph}}> &=& \frac{3\pi I_{\rm{ph}}}{16} \sin^2 \theta_{\rm{obs}} \cos^2 \theta_{\rm{disk}} \sin \theta_{\rm{disk}} \label{eq:Q_ph}\\
    <U_{\rm{ph}}> &=& 0. \label{eq:U_ph}
\end{eqnarray}

\subsubsection{Radiation from the photosphere in the radial direction of the CSM disk ($\tau_{\rm{csm,r}} (r_{\rm{sh}}) < 1$)}

In cases with $\tau_{\rm{csm,r}}(r_{\rm{sh}}) < 1$, where there is a photosphere in the shocked region, we assume that unpolarized photons emerge from the shocked region with $L_{\rm{ph}}=L_{\rm{sh}}$ and a part of these photons are scattered with the probability proportional to $\tau_{\rm{csm,r}}(r_{\rm{sh}})$. Thus, we calculate $<Q_{\rm{ph}}>$ and $<U_{\rm{ph}}>$ using the same formulas as in the previous subsection, but multiply them by $\tau_{\rm{csm,r}}(r_{\rm{sh}})$.

\subsubsection{Radiation from the photosphere at the surface of the CSM disk}

The polarization degree of the radiation from the photosphere at the surface of the CSM disk depends on the exact density distribution of the surface of the disk. In this paper, we assume that the boundary of the CSM disk is sharply defined by a sudden density drop from $\rho_{\rm{disk}}$ to zero, and that the radiation from the disk surface is unpolarized, for simplicity.
Thus, we get the following formulas:
\begin{eqnarray}
    <I_{\rm{surface}}> &=& \frac{L_{\rm{surface}}}{4 \pi (1-\sin \theta_{\rm{disk}})}, \label{eq:I_surface}\\
    <Q_{\rm{surface}}> &=& 0, \label{eq:Q_surface}\\
    <U_{\rm{surface}}> &=& 0. \label{eq:U_surface}
\end{eqnarray}

We note that, in reality, the density distribution at the disk boundary should not be sharply defined and thus the radiation from the disk surface can be polarized to some extent.

\section{Results} \label{sec:results}

\subsection{Polarization properties}

The obtained polarization angle remains constant at zero over time, where we define the polarization angle as measured from the $z$-axis. This means that the polarization angle is always aligned with the polar axis of the CSM disk on the plane of the sky in our configuration (see Figures~\ref{fig:fig1} and \ref{fig:fig2}). Since we assumed that the radiation from the surface of the CSM disk is unpolarized (see Equations~\ref{eq:I_surface}-\ref{eq:U_surface}), it can only reduce the net polarization degree by adding unpolarized radiation and cannot affect the polarization angle. The only source of polarization is the radiation from the photosphere, where $<Q_{\rm{ph}}>$ is always positive and $<U_{\rm{ph}}>$ is always zero (see Equations~\ref{eq:I_ph}-\ref{eq:U_ph}). As a result, the polarization angle (Equation~\ref{eq:pol_angle}) remains fixed at zero degrees ($\theta_{P}=0$). This outcome is consistent with expectations, as the spatial relationship between the observer and the scattering bodies remains constant over time.

\begin{figure}
   \centering
   \includegraphics[width=\hsize]{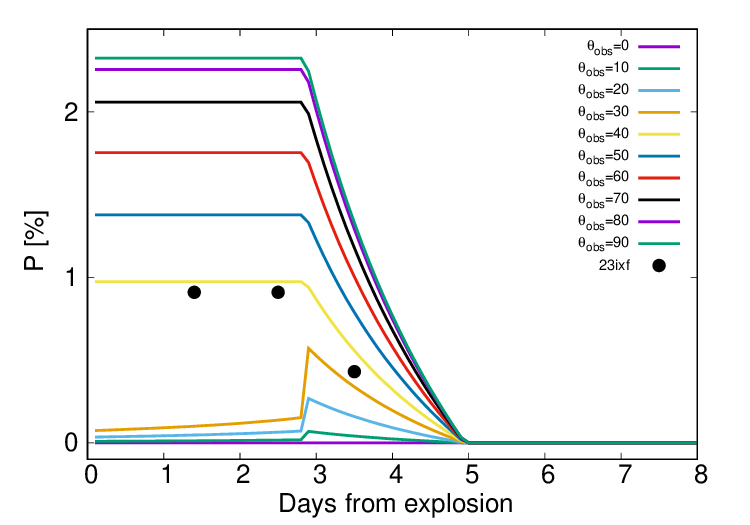}
      \caption{Time volution of the polarization degree. The cases with the parameters of $\theta_{\rm{disk}}=60$ degrees, $s=2.0$, $M_{\rm{csm}}=2\times10^{-3}$ M$_{\odot}$, and $r_{\rm{out}}=3\times 10^{14}$ cm, but with various values of $\theta_{\rm{obs}}$ are shown. The black points show the observed polarization in SN 2023ixf (see Section~\ref{sec:characteristic_value}).  
              }
         \label{fig:fig4}
   \end{figure}

Figure~\ref{fig:fig4} shows the time volution of the polarization degree for the cases with the fiducial parameters of $\theta_{\rm{disk}}=60$ degrees, $s=2.0$, $M_{\rm{csm}}=2\times10^{-3}$ M$_{\odot}$, and $r_{\rm{out}}=3\times 10^{14}$ cm, but with different values of $\theta_{\rm{obs}}$. Larger values of $\theta_{\rm{obs}}$ (i.e., viewing angles closer to the equatorial plane) produce greater degrees of polarization. As the viewing angle decreases (i.e., viewing angles closer to the pole of the CSM disk), the spatial distribution of scattered photons becomes more circularly symmetric, resulting in a reduced net polarization due to the cancellation of the linear polarization components. The decline and eventual disappearance of the polarization signals correspond to the times when the optical depth in front of the interaction shock drops below unity ($\tau_{\rm{csm},r} (r_{\rm{sh}})=1$), and when the interaction shock reaches the outer edge of the confined CSM, respectively. Thus, these times do not depend on the viewing angle.
%
In the cases where the observer see the surface of the CSM disk ($\theta_{\rm{obs}} \leq \pi/2-\theta_{\rm{disk}}$), the net polarization is suppressed by the unpolarized radiation from the surface of the CSM disk, until the CSM becomes optically thin. Thus, we see gradual increase of polarization until $\sim 3$ days after the explosion in the cases with $\theta_{\rm{obs}} \leq 30$ in Figure~\ref{fig:fig4}.
%
%
The observed polarizaion of SN~2023ixf would prefer $\theta_{\rm{obs}}=40$ degrees for the assumed values of the other parameters.

\begin{figure}
   \centering
   \includegraphics[width=\hsize]{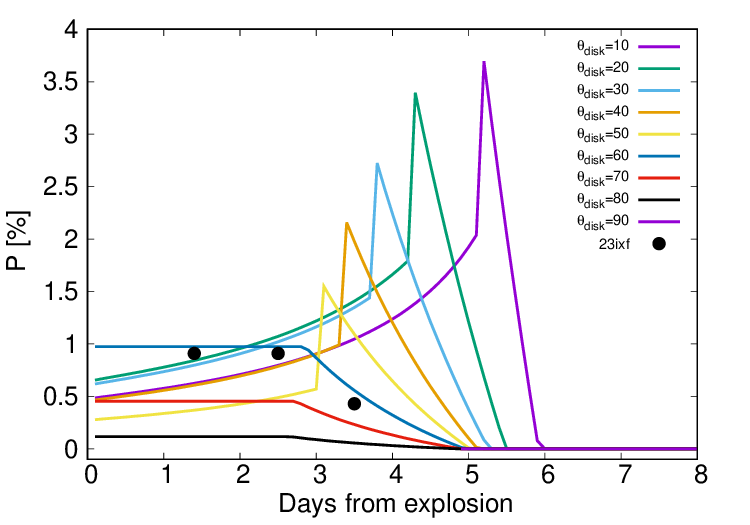}
      \caption{Time volution of the polarization degree. The cases with the parameters of $\theta_{\rm{obs}}=40$ degrees, $s=2.0$, $M_{\rm{csm}}=2\times10^{-3}$ M$_{\odot}$, and $r_{\rm{out}}=3\times 10^{14}$ cm, but with various values of $\theta_{\rm{disk}}$ are shown. The black points are the same as in Figure~\ref{fig:fig4}.  
              }
         \label{fig:fig5}
   \end{figure}

Figure~\ref{fig:fig5} shows the time volution of the polarization degree for the cases with the fiducial parameters of $\theta_{\rm{obs}}=40$ degrees, $s=2.0$, $M_{\rm{csm}}=2\times10^{-3}$ M$_{\odot}$, and $r_{\rm{out}}=3\times 10^{14}$ cm, but with different values of $\theta_{\rm{disk}}$. For smaller opening angle of the CSM disk, the peak polarization degree is larger. This is because the average scattering angles become closer to 90 degrees, whose scattering creates the largest polarization degree, for smaller values of $\theta_{\rm{disk}}$.
There are also polarization rises, as in Figure~\ref{fig:fig4}, for the cases with $\theta_{\rm{obs}} \leq \pi/2-\theta_{\rm{disk}}$ (i.e., $\theta_{\rm{disk}} \leq 50$). For cases with smaller $\theta_{\rm{disk}}$, this polarization rise is more substantial. This is because the unpolarized radiation escaping from the surface of the CSM disk is larger compared to the polarized radiation originating from the photosphere (i.e., $L_{\rm{surface}}/L_{\rm{ph}}$ is larger) for cases with smaller $\theta_{\rm{disk}}$.
The timing of the disappearance of the polarization is slightly earlier in cases with larger $\theta_{\rm{disk}}$. This is because the density of the CSM is smaller and thus the shock evolves faster, in cases with larger $\theta_{\rm{disk}}$.
The observed polarizaion of SN~2023ixf would prefer $\theta_{\rm{disk}}=60$ degrees for the assumed values of the other parameters.

\begin{figure}
   \centering
   \includegraphics[width=\hsize]{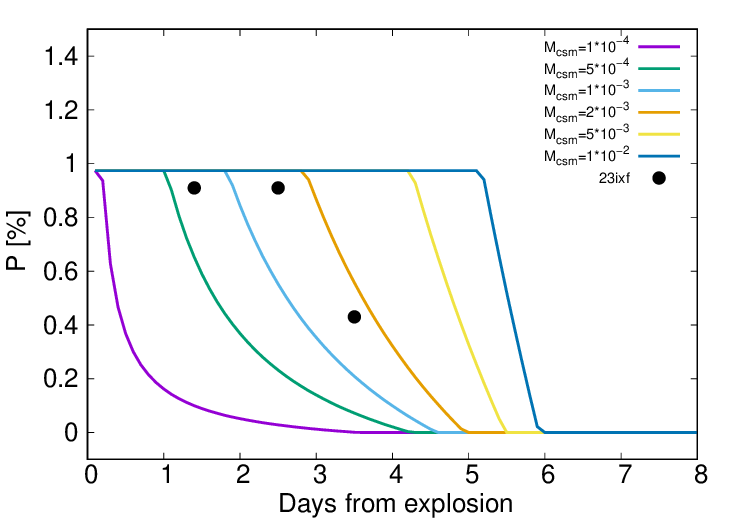}
      \caption{Time volution of the polarization degree. The cases with the parameters of $\theta_{\rm{obs}}=40$ degrees, $\theta_{\rm{disk}}=60$ degrees, $s=2.0$, and $r_{\rm{out}}=3\times 10^{14}$ cm, but with various values of $M_{\rm{csm}}$ are shown. The black points are the same as in Figure~\ref{fig:fig4}.
              }
         \label{fig:fig6}
   \end{figure}

Figure~\ref{fig:fig6} shows the time volution of the polarization degree for the cases with the fiducial parameters of $\theta_{\rm{obs}}=40$ degrees, $\theta_{\rm{disk}}=60$ degrees, $s=2.0$, and $r_{\rm{out}}=3\times 10^{14}$ cm, but with different values of $M_{\rm{csm}}$.
Since we observe the system through the unshocked CSM for these assumed parameters, the maximum polarization degree, which is determined by the geometry between the observer and the scattering bodies, does not change for all the cases here. On the other hand, the timing of the disappearance of the polarization largely depends on the value of the CSM mass. This is because the shock evolution is slower in cases with higher values of $M_{\rm{csm}}$.
Since the location of the interaction shock when the unshocked CSM becomes optically thin is closer to the outer edge of the CSM disk, the decline of polarization is faster for cases with larger $M_{\rm{csm}}$.
The observed polarizaion of SN~2023ixf would prefer $M_{\rm{csm}}=10^{-3}$ M$_{\odot}$ for the assumed values of the other parameters.

\begin{figure}
   \centering
   \includegraphics[width=\hsize]{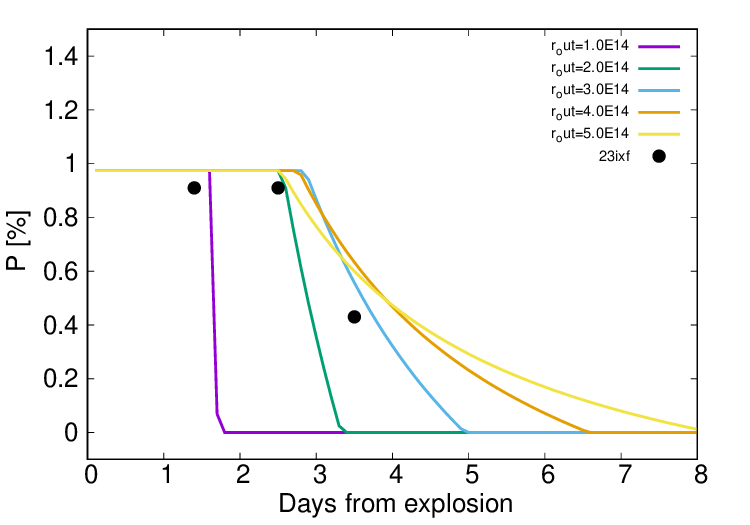}
      \caption{Time volution of the polarization degree. The cases with the parameters of $\theta_{\rm{obs}}=40$ degrees, $\theta_{\rm{disk}}=60$ degrees, $M_{\rm{csm}}=2\times10^{-3}$ M$_{\odot}$, and $s=2.0$, but with various values of $r_{\rm{out}}$ are shown. The black points are the same as in Figure~\ref{fig:fig4}.
              }
         \label{fig:fig7}
   \end{figure}

Figure~\ref{fig:fig7} shows the time volution of the polarization degree for the cases with the fiducial parameters of $\theta_{\rm{obs}}=40$ degrees, $\theta_{\rm{disk}}=60$ degrees, $M_{\rm{csm}}=2\times10^{-3}$ M$_{\odot}$, and $s=2.0$, but with different values of $r_{\rm{out}}$.
Since the shock evolution is identical for all the cases here, the different values of $r_{\rm{out}}$ produce different times of the disappearance of the polarization. On the other hand, the location where the unshocked CSM becomes optically thin does not strongly depend on the outer radius of the CSM disk ($\sim 2\times 10^{14}$ cm for the assumed density distribution), creating similar times of the polarization decline.
The observed polarizaion of SN~2023ixf would prefer $r_{\rm{out}}=2\times 10^{14}$ cm for the assumed values of the other parameters.

\begin{figure}
   \centering
   \includegraphics[width=\hsize]{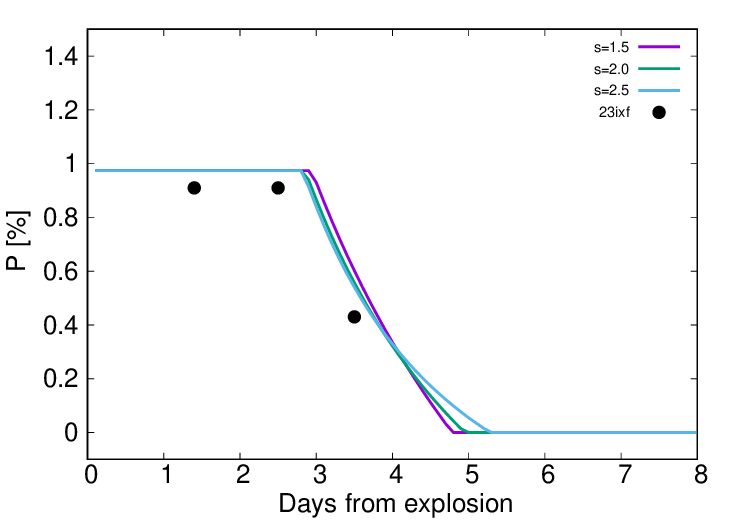}
      \caption{Time volution of the polarization degree. The cases with the parameters of $\theta_{\rm{obs}}=40$ degrees, $\theta_{\rm{disk}}=60$ degrees, $M_{\rm{csm}}=2\times10^{-3}$ M$_{\odot}$, and $r_{\rm{out}}=3\times 10^{14}$ cm, but with various values of $s$ are shown. The black points are the same as in Figure~\ref{fig:fig4}.
              }
         \label{fig:fig8}
   \end{figure}

Figure~\ref{fig:fig8} shows the time volution of the polarization degree for the cases with the fiducial parameters of $\theta_{\rm{obs}}=40$ degrees, $\theta_{\rm{disk}}=60$ degrees, $M_{\rm{csm}}=2\times10^{-3}$ M$_{\odot}$, and $r_{\rm{out}}=3\times 10^{14}$ cm, but with different values of $s$. These models do not produce much difference. Since we assume the single-scattering approximation, the distribution of the scattering bodies does not affect the polarization degree. Although the difference of $s$ can affect the evolution of the interaction shock, the difference of the results is not much for the assumed values of the parameters.

\subsection{Characteristic values of the polarization and CSM parameter estimation} \label{sec:characteristic_value}

   \begin{figure*}
   \centering
   \includegraphics[width=0.49\hsize]{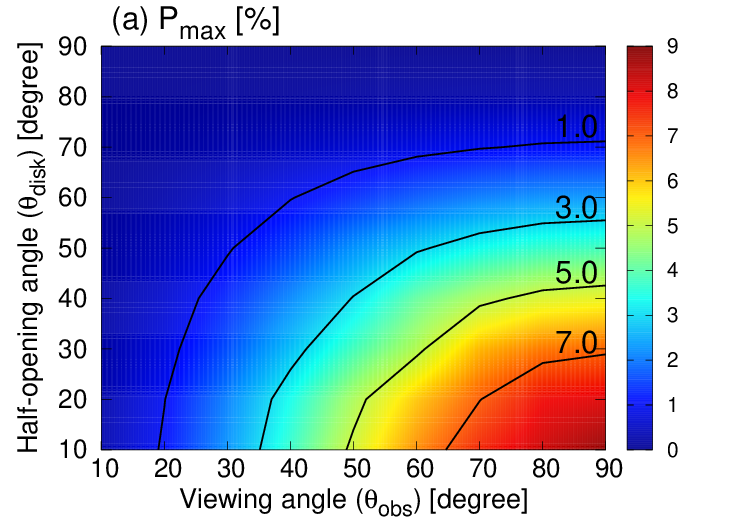}
   \includegraphics[width=0.49\hsize]{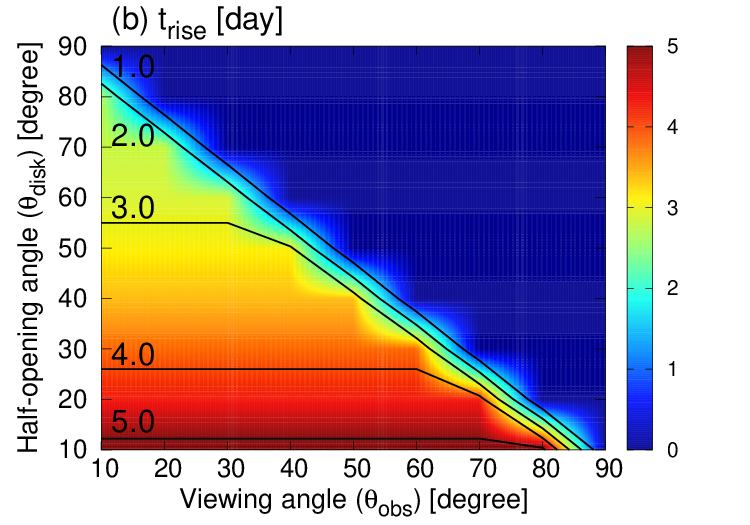}
      \caption{Parameter dependence of the maximum polarization degree ($P_{\rm{max}}$) and the rise time ($t_{\rm{rise}}$). (a) $P_{\rm{max}}$ for cases with $M_{\rm{csm}}= 2\times10^{-3}$ M$_{\odot}$, $r_{\rm{out}}=3\times 10^{14}$ cm, and $s=2$, but with various values of $\theta_{\rm{obs}}$ and $\theta_{\rm{disk}}$. (b) Same as the panel (a), but for $t_{\rm{rise}}$. The region with $\theta_{\rm{obs}} \geq \pi/2-\theta_{\rm{disk}}$ (i.e., cases with edge-on views) corresponds to $t_{\rm{rise}}=0$. The wavy pattern around $\theta_{\rm{obs}} = \pi/2 - \theta_{\rm{disk}}$ is an artifact caused by the limited size of the computational grid.
              }
         \label{fig:fig9}
   \end{figure*}

      \begin{figure*}
   \centering
   \includegraphics[width=0.49\hsize]{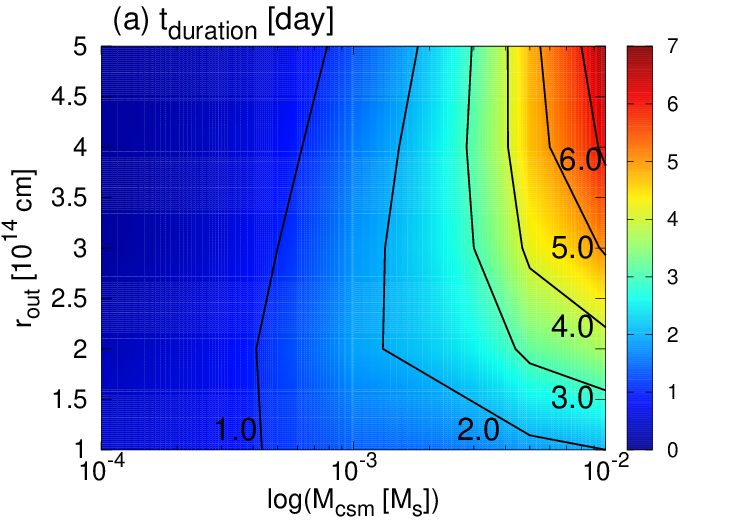}
   \includegraphics[width=0.49\hsize]{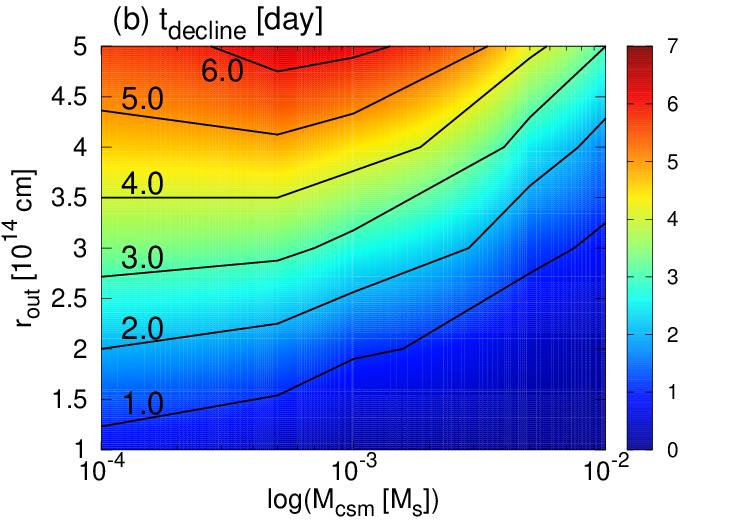}
      \caption{Parameter dependence of the duration ($t_{\rm{duration}}$) and decline time ($t_{\rm{decline}}$) of the polarization. (a) $t_{\rm{duration}}$ for cases with $\theta_{\rm{obs}}=40$ degrees, $\theta_{\rm{disk}}=60$ degrees, and $s=2$, but with various values of $M_{\rm{csm}}$ and $r_{\rm{out}}$. (b) Same as the panel (a), but for $t_{\rm{decline}}$.
              }
         \label{fig:fig10}
   \end{figure*}

      \begin{figure*}
   \centering
   \includegraphics[width=0.49\hsize]{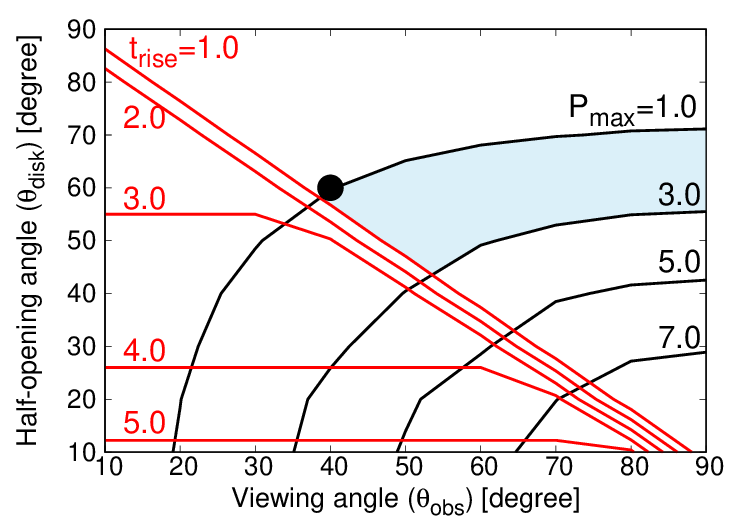}
   \includegraphics[width=0.49\hsize]{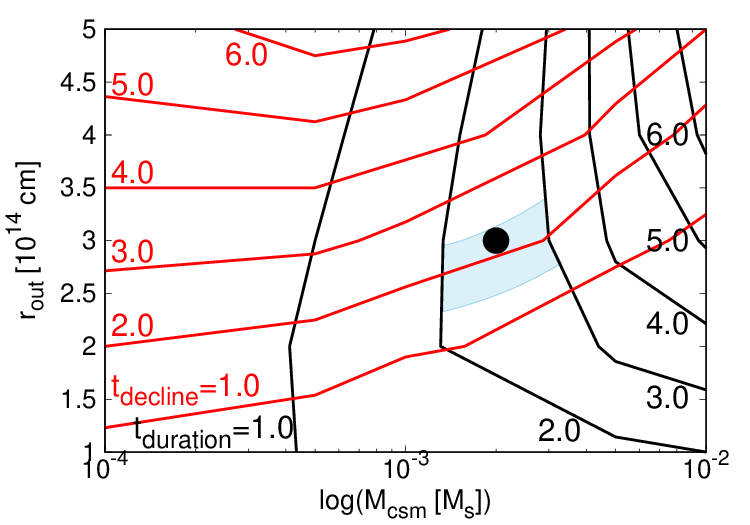}
      \caption{Application of the observed polarization of the Type~II SN~2023ixf to the models in Figures~\ref{fig:fig9}~and~\ref{fig:fig10}. The black dots show our adopted good-match values, while the blue regions correspond to the regions with potential good matches.
              }
         \label{fig:fig11}
   \end{figure*}

As shown in the previous subsection, the temporal evolution of polarization in Type II SNe with confined CSM can be used to roughly estimate the CSM parameters. As characteristic parameters for the temporal evolution of polarization, we consider the following quantities: the polarization maximum ($P_{\rm{max}}$; the maximum polarization degree), the rise time ($t_{\rm{rise}}$; the time from the explosion until the polarization reaches its maximum), the duration ($t_{\rm{duration}}$; the time from the explosion until the polarization begins to decline), the decline time ($t_{\rm{decline}}$; the time from the onset of the polarization decline to its disappearance). Here, in cases without a polarization rise (i.e., a constant polarization at early phases), we set $t_{\rm{rise}}$ as zero.

The maximum polarization degrees are primarily determined by the values of $\theta_{\rm{obs}}$ and $\theta_{\rm{disk}}$, rather than by those of $M_{\rm{csm}}$, $r_{\rm{out}}$ and $s$. Since the maximum polarization is determined by the average scattering angles of the photons emerging from the photosphere, the geometry between the photosphere and the observer (related to $\theta_{\rm{obs}}$) and the shape of the photosphere (related to $\theta_{\rm{disk}}$) are the important factors. Figure\ref{fig:fig9}a shows the maximum polarization degrees for models with varying values of $\theta_{\rm{obs}}$ and $\theta_{\rm{disk}}$, assuming $M_{\rm{csm}}=2\times10^{-3}$ M$_{\odot}$, $r_{\rm{out}}=3\times 10^{14}$ [cm], and $s=2$. Smaller opening angles and larger viewing angles result in higher maximum polarization degrees.
At the same time, the presence of the polarization rise would also provide a constraint for the values of $\theta_{\rm{obs}}$ and $\theta_{\rm{disk}}$. If there is a rise, we can get the constraint of $\theta_{\rm{obs}} \leq \pi/2-\theta_{\rm{disk}}$. Otherwise, the viewing angle should be larger than $\pi/2-\theta_{\rm{disk}}$. Figure~\ref{fig:fig9}b shows the rise time of the polarization, which is strongly correlated with $\theta_{\rm{disk}}$ and thus can be used for estimating its value.

The duration and decline time of the polarization are more sensitive to the values of $M_{\rm{csm}}$ and $r_{\rm{out}}$ than to those of $\theta_{\rm{obs}}$, $\theta_{\rm{disk}}$, or $s$. Figure~\ref{fig:fig10}a shows the duration and decline time for models with varying values of $M_{\rm{csm}}$ and $r_{\rm{out}}$, assuming $\theta_{\rm{obs}}=40$ degrees, $\theta_{\rm{disk}}=60$ degrees, and $s=2$. The duration is more strongly correlated with the value of $M_{\rm{csm}}$ than with that of $r_{\rm{out}}$, whereas the decline time shows the opposite trend. Therefore, by measuring the duration and decline time with polarimetric observations, we can constrain the values of $M_{\rm{csm}}$ and $r_{\rm{out}}$.

Here, we consider applying these results to observations of the Type II SN~2023ixf, which is the only SN that has early-time polarimetric observations with a good temporal coverage, in order to derive its CSM parameters. SN~2023ixf shows polarization with time varying degrees and angles \citep[][]{Vasylyev2023, Vasylyev2025}. Following \citet[][]{Vasylyev2023, Vasylyev2025}, we interpret the first component of the observed polarization - with polarization angles of $\sim 160$ degrees observed until $\sim 4$ days - as originating from the confined CSM, which we have modeled in this study. For SN~2023ixf, we roughly estimate the values of the characteristic parameters as follows: $P_{\rm{max}}\sim 0.9$ \%, $t_{\rm{rise}} \lesssim 1$ day, $t_{\rm{duration}} \sim 2.5 \pm 0.5$ days, $t_{\rm{decline}} \sim 2 \pm 0.5$ days.

Based on the values of $P_{\rm{max}}$ and $t_{\rm{rise}}$, we estimate good-fit values of $\theta_{\rm{obs}}$ and $\theta_{\rm{disk}}$ as $\gtrsim 40$ degrees and $\sim 60$ degrees, respectively, using the relations in Figure~\ref{fig:fig9} (see Figure~\ref{fig:fig11}a). We note that, as we will discuss in the next section, the polarization degrees obtained from our calculations may be overestimated due to the single-scattering assumption (see Section~\ref{sec:limitation} for the details). Therefore, the actual value of $\theta_{\rm{disk}}$ might be slightly smaller. If we take the region with $1 \leq P_{\rm{max}} \leq 3$ \% and  $t_{\rm{rise}} \leq 1$ day, the good-fit values would be $\theta_{\rm{obs}} \lesssim 40$ degrees and $50 \lesssim \theta_{\rm{disk}} \lesssim 60$ degrees (see the blue hatched region in Figure~\ref{fig:fig11}a). Based on the values of $t_{\rm{duration}}$ and $t_{\rm{decline}}$, we estimate good-fit values of $M_{\rm{csm}}$ and $r_{\rm{out}}$ as $\sim 2 \times 10^{-3}$ M$_{\odot}$ and $\sim 3\times 10^{14}$ cm, respectively, using the relations in Figure~\ref{fig:fig10} (see Figure~\ref{fig:fig11}b).

The observed polarization angle of $\sim 160$ degrees suggests that the polar axis of the confined CSM disk of SN~2023ixf, if it has disk-like distribution, is tilted by $\sim 160$ degrees from the north-south direction on the night sky. Interestingly, this orientation correlates with the axis of the asymmetric structure of the SN~2023ixf explosion \citep[][]{Vasylyev2025}. This alignment may indicate that the mechanisms creating the aspherical structures of both the confined CSM and the SN ejecta are related. Thus, the formation of the confined CSM might be driven by processes intrinsic to the progenitor star itself, rather than by interactions with a companion star, unless all the axes of the explosion, the CSM disk and the companion orbit are aligned. This should be investigated with more observations of early polarimetry of Type II SNe.

\section{Limitations of the calculations} \label{sec:limitation}

In this section, we discuss the limitations of our methodology to predict the polarization signals from confined CSM in Type II SNe. Our calculation method has several simplifications. The biggest one is the assumption of the single scattering (see Section~\ref{sec:calculations}). In our calculations, we assumed that the location of the photosphere corresponds to $\tau_{\rm{csm},r} (r_{\rm{ph}})=1$, and that the photons emerging from the photosphere are scattered only once, as the atmosphere above it is optically thin. However, they can in principle get multiple scatterings, which reduce the polarization degree. This would depend on the configuration of scattering bodies, i.e., on where the last scatterings typically happen.

\begin{figure*}
   \centering
   \includegraphics[width=\hsize]{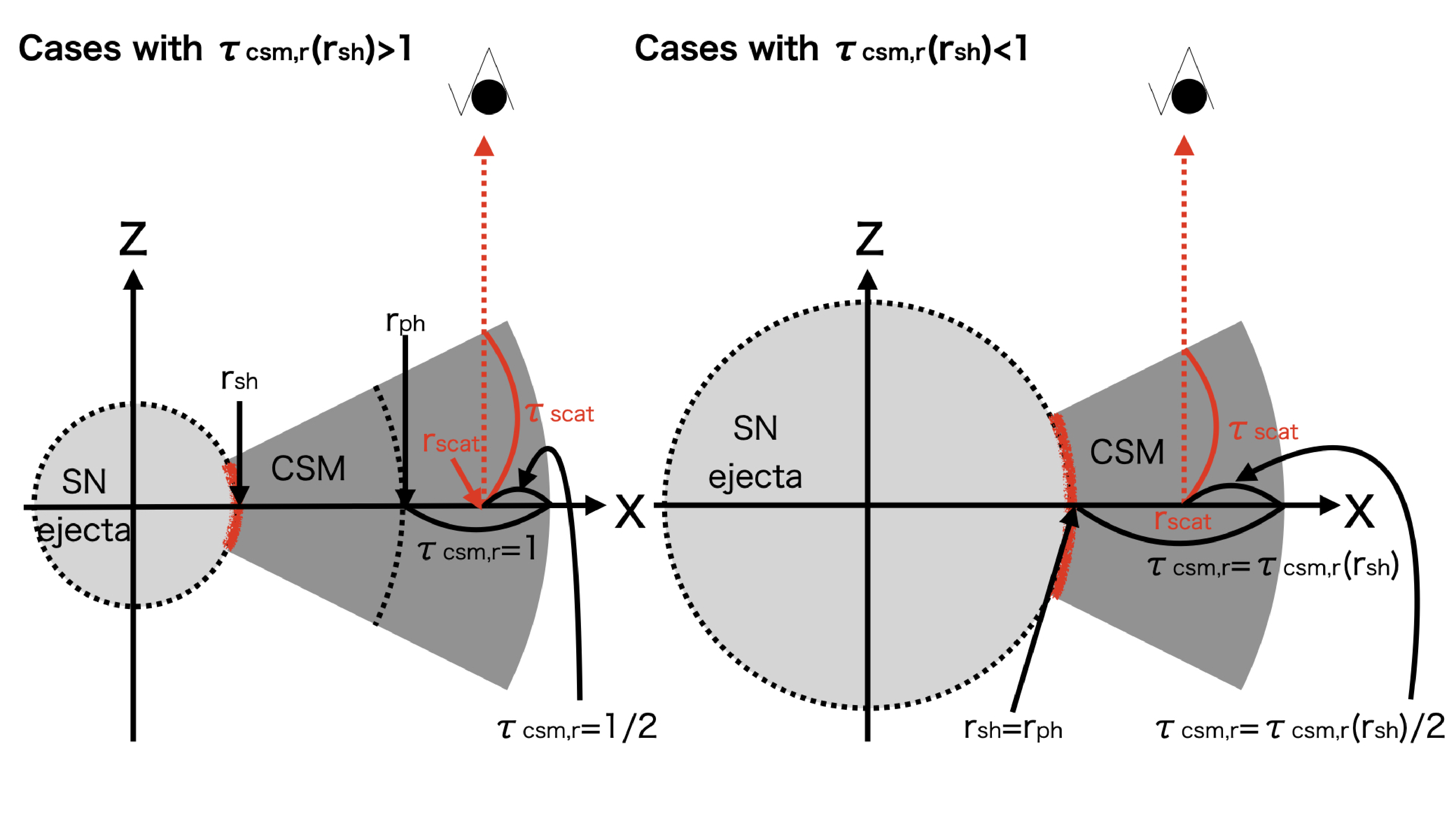}
      \caption{Schematic illustration of our calculations for $\tau_{\rm{scat}}$.
              }
         \label{fig:fig12}
   \end{figure*}

For roughly estimating the probability of the multiple scatterings, we check the average locations of the scattering events of the photons originating from the photosphere ($r_{\rm{scat}}$) and assess the optical depth from the places to the observers ($\tau_{\rm{scat}}$), in the single scattering situation (see Figure~\ref{fig:fig12}). If this optical depth is larger than unity, the scattered photons can have additional scattering events, and thus the calculated polarization degrees can be smaller than those in Section~\ref{sec:results}.
When $\tau_{\rm{csm,r}}(r_{\rm{sh}}) \geq 1$, the average radius of the scatterings is estimated by $\tau_{\rm{csm,r}}(r_{\rm{scat}})=1/2$. Thus, we get
\begin{equation}
    r_{\rm{scat}} = \left( r_{\rm{out}}^{-(s-1)} + \frac{s-1}{2\kappa_{\rm{es}}D} \right)^{-1/(s-1)}.
\end{equation}
Thus, 
\begin{equation}
    \tau_{\rm{csm,h}} (r_{\rm{scat}}) = \theta_{\rm{disk}} \left( \kappa_{\rm{es}} D r_{\rm{out}}^{-(s-1)} + \frac{s-1}{2} \right).
\end{equation}
For simplicity, we assume the optical depth from the average locations of the scattering events to the observer wtih $\theta_{\rm{obs}}$ as follows:
\begin{equation}
    \tau_{\rm{scat}} (\theta_{\rm{obs}}) = (1-\cos \theta_{\rm{obs}}) \tau_{\rm{csm,r}} (r_{\rm{scat}}) + \cos \theta_{\rm{obs}} \tau_{\rm{csm,h}} (r_{\rm{scat}}).
\end{equation}

When $\tau_{\rm{csm,r}}(r_{\rm{sh}}) < 1$, we define $r_{\rm{scat}}$ using $\tau_{\rm{csm,r}}(r_{\rm{scat}})=1/2 \times \tau_{\rm{csm,r}}(r_{\rm{sh}})$ as follows:
\begin{equation}
    r_{\rm{scat}} = \left( \frac{1}{2} r_{\rm{sh}}^{-(s-1)} + \frac{1}{2} r_{\rm{out}}^{-(s-1)} \right)^{-1/(s-1)}.
\end{equation}
Thus,
\begin{equation}
    \tau_{\rm{csm,h}} (r_{\rm{scat}}) = \frac{\kappa_{\rm{es}} D \theta_{\rm{disk}}}{2} \left(r_{\rm{sh}}^{-(s-1)} +  r_{\rm{out}}^{-(s-1)} \right),
\end{equation}
and,
\begin{eqnarray}
    \tau_{\rm{scat}} (\theta_{\rm{obs}}) &=& \kappa_{\rm{es}} D \left( \frac{1-\cos \theta_{\rm{obs}}}{2(s-1)} + \frac{\theta_{\rm{disk}} \cos \theta_{\rm{obs}}}{2} \right) r_{\rm{sh}}^{-(s-1)} \nonumber\\
    &&+ \kappa_{\rm{es}} D \left( -\frac{1-\cos \theta_{\rm{obs}}}{2(s-1)} + \frac{\theta_{\rm{disk}} \cos \theta_{\rm{obs}}}{2} \right) r_{\rm{out}}^{-(s-1)}.
\end{eqnarray}

\begin{figure}
   \centering
   \includegraphics[width=\hsize]{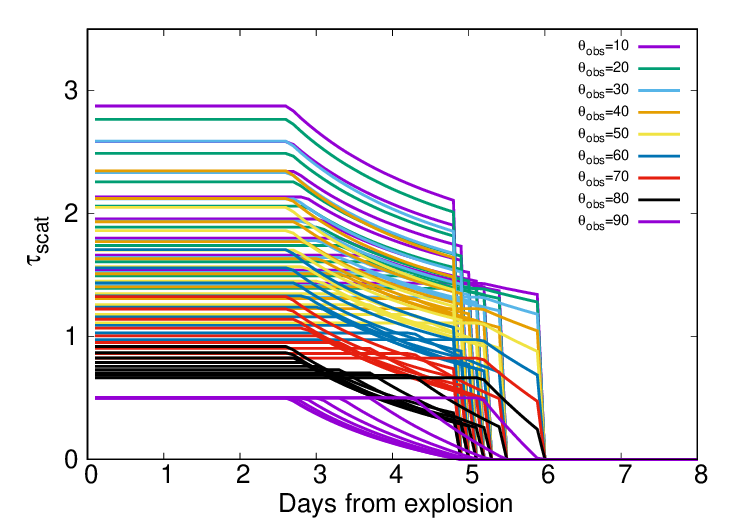}
      \caption{Time evolution of $\tau_{\rm{scat}}$ in cases with different values of $\theta_{\rm{obs}}$ and $\theta_{\rm{disk}}$, with $M_{\rm{csm}}=10^{-3}$ M$_{\odot}$, $r_{\rm{out}}=2\times 10^{14}$ [cm], and $s=2$.
              }
         \label{fig:fig13}
   \end{figure}

\begin{figure}
   \centering
   \includegraphics[width=\hsize]{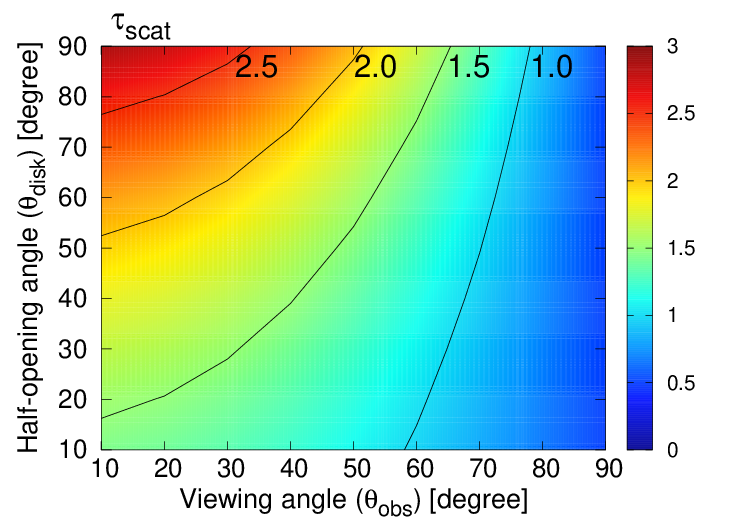}
      \caption{Maximum $\tau_{\rm{scat}}$ for cases with different values of $\theta_{\rm{obs}}$ and $\theta_{\rm{disk}}$, with $M_{\rm{csm}}=10^{-3}$ M$_{\odot}$, $r_{\rm{out}}=2\times 10^{14}$ [cm], and $s=2$.
              }
         \label{fig:fig14}
   \end{figure}

Figure~\ref{fig:fig13} shows the time evolution of the values of $\tau_{\rm{scat}}$ for different values of $\theta_{\rm{obs}}$ and $\theta_{\rm{disk}}$. When there is a photosphere (i.e., $\tau_{\rm{csm,r}}(r_{\rm{sh}}) \geq 1$), the locations of the scattering events are constant with time, and thus the values of $\tau_{\rm{scat}}$ are also constant. Once the shocked shell reach to the location with $\tau_{\rm{csm,r}}=1$, the optical depths at the scattering locations decrease with time until the shock reach to the outer edge of the confined CSM. When the value of $\tau_{\rm{scat}}$ is roughly greater than unity, the calculated polarization degrees in the single scattering configuration would be overestimated.
The maximum values of $\tau_{\rm{scat}}$ for different values of $\theta_{\rm{obs}}$ and $\theta_{\rm{disk}}$ are shown in Figure~\ref{fig:fig14}. Cases with smaller viewing angles and larger half-opening angles of the CSM disk have significantly higher values of $\tau_{\rm{scat}}$ than unity. Thus, the calculated polarization degrees for such cases should in reality be reduced by the multiple scatterings. On the other hand, cases with larger calculated polarization degrees (i.e., with larger viewing angles and smaller half-opening angles) are less affected by this multiple scatterings (see also Figure~\ref{fig:fig9}).

The good-fit case for SN~2023ixf (with $\theta_{\rm{obs}}=40$ degrees, $\theta_{\rm{disk}}=60$ degrees, $M_{\rm{csm}}=10^{-3}$ M$_{\odot}$, $r_{\rm{out}}=2\times 10^{14}$ cm, $s=2.0$) has $\tau_{\rm{scat}} \sim 2$. Therefore, taking the multiple-scattering effects into account, the realistic values of the parameters for the confined CSM in SN~2023ixf would be modified with a slightly larger viewing angle and/or a slightly smaller half-opening angle.
We note that this is a very rough estimate of the possibility of multiple-scattering effects, ignoring the realistic distribution of the CSM, and thus we need to conduct multi-dimensional radiative transfer calculations for obtaining proper values of the polarization degrees.

Another simplification that might change the results is the polarization degree of the radiation originating from the surface of the CSM disk, which we assume as zero. In reality, this radiation can have a non-negligible amount of polarization, and might change the polarization degrees at early phases (i.e., when $\tau{\rm{csm,r}} (r_{\rm{sh}}) > 1$) for cases with $\theta_{\rm{obs}} \leq \pi/2-\theta_{\rm{disk}}$. We note that this modification does not change the maximum polarization degrees, the timing of the polarization rises, or the duration of the polarization.

In addition, since we assume that the main radiation source is the interaction between the SN ejecta and the CSM, ignoring the contribution from the SN ejecta, the results from the cases with smaller amount of CSM might be less reliable.

\section{Conclusion} \label{sec:conclusion}

We modeled the polarization signals produced by electron scattering in a Type II SN with a confined, disk-like CSM. 
In this system, the calculated polarization angle remains constant at zero degrees, which is parallel to the axis of the CSM disk, regardless of the parameters of the confined CSM. The temporal evolution of the polarization degree varies depending on the parameters of the confined CSM with a timescale of $\lesssim 10$ days. The general behavior is as follows: Initially, the polarization degree remains constant or gradually increases to several percent or less while the unshocked CSM in front of the interaction shock remains optically thick. It then reaches a peak as the unshocked CSM becomes optically thin. Finally, the polarization declines and eventually vanishes as the interaction shock reaches the outer edge of the CSM disk.
The results demonstrate that the early-time polarization in Type~II SNe can effectively probe the geometry and properties of confined CSM. In particular, the maximum degree and the rise time of the polarization can constrain the viewing angle and the opening angle of the CSM disk, while the duration and the decline time can constrain the mass and extension of the CSM disk. 

As a concrete application to observations, we apply the models to the observed polarization of the Type~II SN 2023ixf and find that it can be explained by a disk-like CSM with the following parameters: the viewing angle of $\theta_{\rm{obs}} \gtrsim 40$ degrees, the half-opening angle of the disk of $\theta_{\rm{disk}} \sim 50-60$ degrees, the CSM mass of $M_{\rm{csm}}\sim 2\times10^{-3}$ M$_{\odot}$, and the outer edge of the CSM disk of $r_{\rm{out}} \sim 3\times 10^{14}$ cm. 
In addition, the alignment between the explosion asymmetry and the CSM disk of SN~2023ixf may suggest a common origin for these asymmetric structures, possibly tied to the progenitor star rather than companion interaction. Early polarimetry of Type II SNe will be essential to test this scenario. Furthermore, the developed method is broadly applicable to various objects with scattering-dominated photospheres, offering a powerful tool for investigating their geometries.

\begin{acknowledgements}
     
We thank Masaomi Tanaka for valuable discussions and insights. The authors wish to acknowledge CSC – IT Center for Science, Finland, for computational resources. This research is supported by the Finnish Ministry of Education and Culture and CSC - IT Centre for Science (Decision diary number OKM/10/524/2022).
T.N. acknowledges support from the Research Council of Finland projects 324504, 328898 and 353019.
K.M. acknowledges support from JSPS KAKENHI grant (JP24KK0070, JP24H01810). 

\end{acknowledgements}

\bibliographystyle{aa} 
\bibliography{aa.bib}




\appendix

\onecolumn

\section{Polarization from a unit area on a photosphere} \label{sec:app1}

    \begin{figure*}
            \includegraphics[width=0.5\hsize]{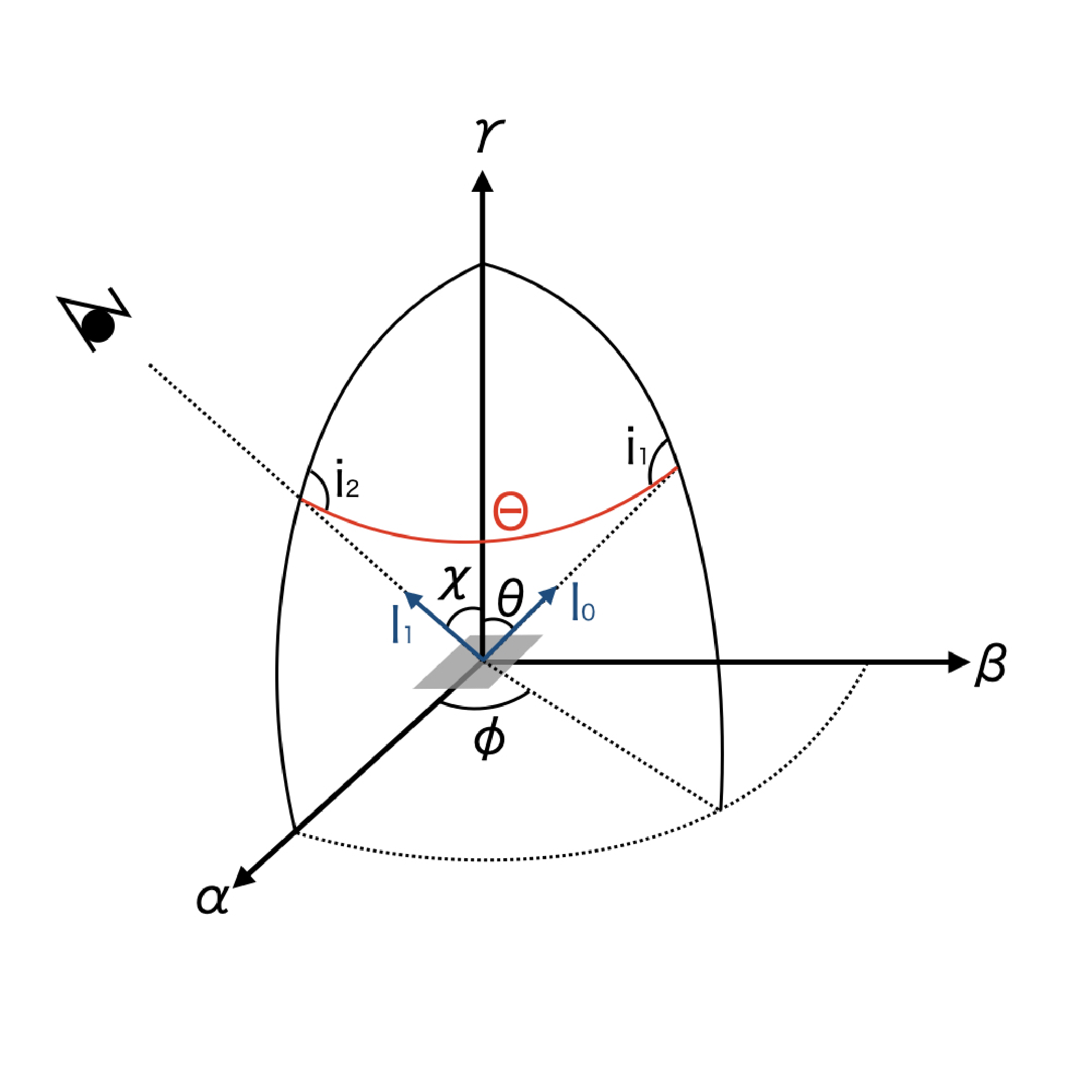}
            \includegraphics[width=0.5\hsize]{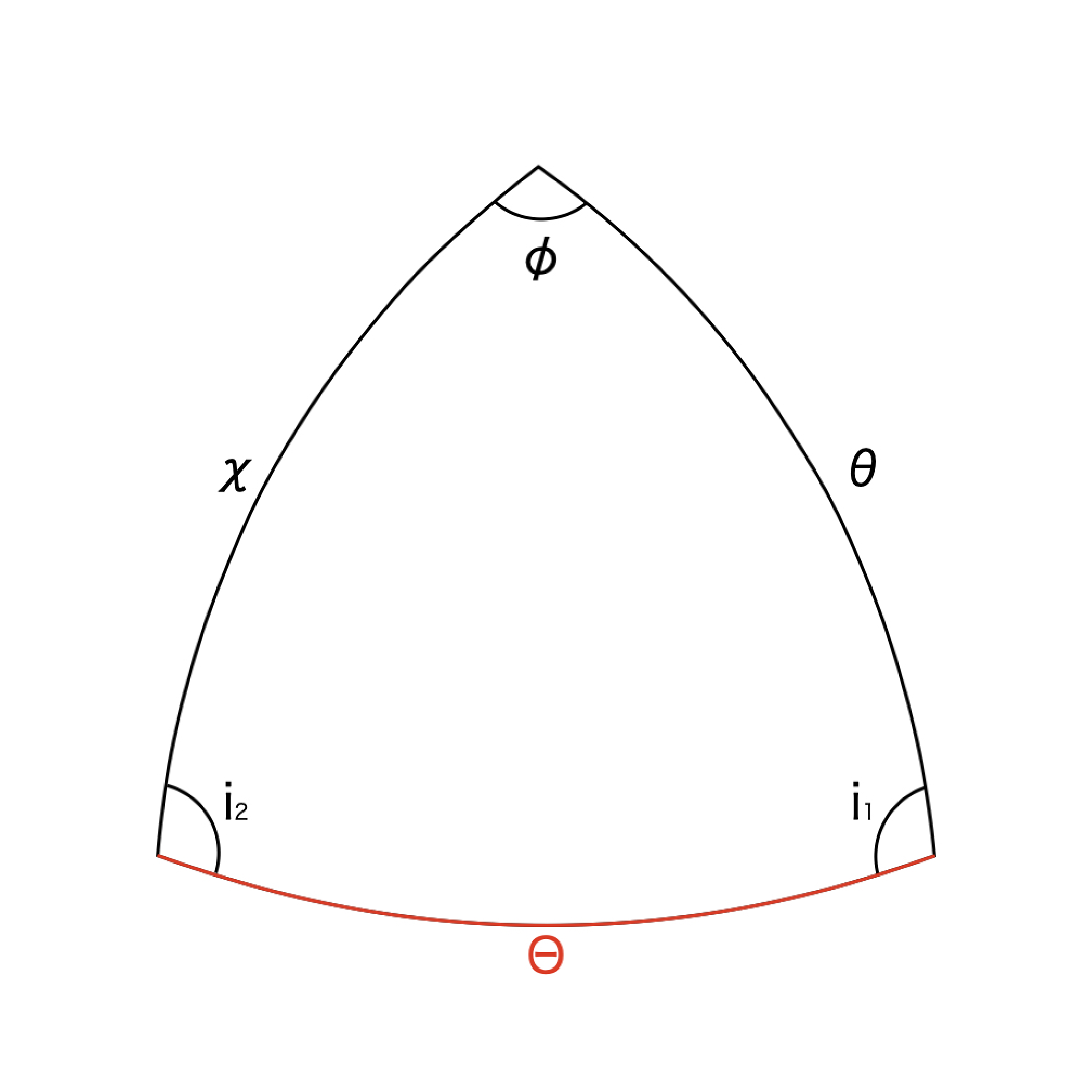}
      \caption{Geometry for scattering. The spherical $\theta$-$\phi$ coordinate is defined in the Cartesian $\alpha$-$\beta$-$\gamma$ coordinate. A photon emerging from a unit area (gray region) at the origin to a direction of ($\theta$, $\phi$) is scattered to the observer's direction ($\chi$,0).
              }
      \label{fig:app1}
   \end{figure*}

We consider a unit area on a ``photosphere" in scattering-dominated gas. In this paper, we define the photosphere as the region where the optical depth, measured from the observer, equals unity. We assume that all photons at the photosphere are unpolarized and are scattered once above the photosphere. We note that possible additional scatterings after the first (i.e., multiple scatterings) are ignored. The Cartesian $\alpha$-$\beta$-$\gamma$ coordinate is defined on a unit area of the photosphere as in Figure~\ref{fig:app1}. We use the spherical $\theta$-$\phi$ coordinate defined in the Cartesian $\alpha$-$\beta$-$\gamma$ coordinate (see Figure~\ref{fig:app1}). The observer is on the $\alpha$-$\gamma$ surface with the viewing angle, $\chi$, from the $gamma$ axis (i.e., ($\theta$,$\phi$)=($\chi$,0)).
The emerging flux at the unit area of the photosphre is expressed as $f_{\rm{in}}$, and its intensity to the perpendicular direction towards the area (i.e., the $\gamma$ direction) as $I_{\rm{in}}$. Since the intensity toward the angle $\theta$ from the $\gamma$ direction ($I_{0}$) is $I_{\rm{in}} \cos \theta$, we get the relation between $f_{\rm{in}}$ and $I_{\rm{in}}$ as follows:
\begin{eqnarray}
    f_{\rm{in}} &=& \int^{\pi/2}_{0} I_0 \mathrm{d} \theta \int^{2\pi}_{0} \mathrm{d} \phi = 2\pi I_{\rm{in}},\\
    \therefore I_{\rm{in}} &=& \frac{f_{\rm{in}}}{2\pi}.
\end{eqnarray}

We consider the photons that are emitted for a unit time to the direction of ($\theta$,$\phi$) from this unit area, whose Stokes parameters are ($\mathbb{S}_{0}=(I_{0},0,0)$). Some of these photons would be scattered to the observer's direction, $(\theta,\phi)=(\chi,0)$, at a certain radius, whose Stokes parameters are denoted as ($\mathbb{S}_{1}=(I_{1},Q_{1},U_{1})$).
We can calculate this conversion of these Stokes vectors of the photons before and after the scatterings, using the scattering matrix ($\mathbb{R} (\Theta)$) and the rotation matrix ($\mathbb{L} (\psi)$), as follows:
\begin{equation}
\mathbb{S}_{1} = \mathbb{L} (\pi-i_{2}) \mathbb{R} (\Theta) \mathbb{L} (-i_{1}) \mathbb{S}_{0},
\end{equation}
where,
\begin{eqnarray}
    \mathbb{R} (\Theta) = \alpha \left(
    \begin{array}{ccc}
    \cos^{2} (\Theta)+1 & \cos^{2} (\Theta)-1 & 0 \\
    \cos^{2} (\Theta)-1 & \cos^{2} (\Theta)+1 & 0 \\
    0 & 0 & 2 \cos \Theta
    \end{array}
\right),
\end{eqnarray}
\begin{eqnarray}
    \mathbb{L} (\psi) = \left(
    \begin{array}{ccc}
    1 & 0 & 0 \\
    0 & \cos 2\psi & \sin 2\psi \\
    0 & -\sin 2\psi & \cos 2\psi
    \end{array}
\right),
\end{eqnarray}
and the $\Theta$ is the angle between the angles before and after the scatterings, and the angles of $i_1$ and $i_2$ are defined as in Figure~\ref{fig:app1}. Based on the spherical trigonometry, we can derive these angles:
\begin{eqnarray}
    \cos \Theta &=& \cos \chi \cos \theta + \sin \chi \sin \theta \cos \phi,\\
    \cos i_{2} &=& \frac{\cos \theta - \cos \chi \cos \Theta}{\sin \chi \sin \Theta}.\\
\end{eqnarray}

Using these values, we obtain the conversion equations.
\begin{equation}
    \begin{cases}
    I_{1} (\theta, \phi) = \alpha I_{0} (1+ \cos^{2} \chi \cos^{2}\theta + 2\cos \chi \sin \chi \cos \theta \sin \theta \cos \phi + \sin^{2} \chi \sin^{2} \theta \cos^{2} \phi),\\
    Q_{1} (\theta, \phi) = \alpha I_{0} (-1+2\sin^{2} \theta + \cos^{2}\chi \cos^{2}\theta + 2\cos\chi \sin\chi \cos\theta \sin\theta \cos\phi + (\sin^{2}\chi -2) \sin^{2}\theta \cos^{2}\phi),\\
    U_{1} (\theta, \phi) = -2 \alpha I_{0} (\sin\chi \cos\theta \sin\theta \sin\phi - \cos\chi \sin^{2}\theta \cos\phi \sin\phi).
    \end{cases}
\end{equation}

Next, we consider all the contribution from all the photons that are emitted to all directions from the unit area during a unit time. We can simply integrate all the contribution to the total Stokes parameters from all the photons that are scattered to the observer's direction, as follows:
\begin{eqnarray}
    I_{\rm{total}} (\chi) &=& \int^{2\pi}_{0} d\phi \int^{\pi/2}_{0} d\theta \left\{ I_{1}(\theta,\phi) \sin \theta \right\},\\
    &=& \alpha I_{\rm{in}} \int^{\pi/2}_{0} d\theta \Bigg\{ 
    (\sin\theta + \cos^{2}\chi \cos^{2}\theta \sin\theta) \cos\theta \left( \int^{2\pi}_{0} d\phi \right) + 2 \cos\chi \sin\chi \cos^{2}\theta \sin^{2}\theta \left( \int^{2\pi}_{0} \cos\phi d\phi \right)\nonumber\\
    && \hspace{10cm}+ \sin^{2}\chi \cos\theta \sin^{3}\theta \left( \int^{2\pi}_{0} \cos^{2} \phi d\phi \right) \Bigg\}, \\
    &=& \alpha \pi I_{\rm{in}} \left\{ 2 \left( \int^{\pi/2}_{0} \cos\theta \sin\theta d\theta \right)+ 2\cos^{2}\chi \left( \int^{\pi/2}_{0} \cos^{3}\theta \sin\theta d\theta \right) + \sin^{2}\chi \left( \int^{\pi/2}_{0} \cos\theta \sin^{3}\theta d\theta \right) \right\},\\
    &=& \alpha \pi I_{\rm{in}} \left( 1 + \frac{1}{2} \cos^{2}\chi + \frac{1}{4} \sin^{2}\chi \right),\\
    &=& \alpha \pi I_{\rm{in}} \left( \frac{3}{2} - \frac{1}{4} \sin^{2}\chi \right).
\end{eqnarray}

\begin{eqnarray}
    Q_{\rm{total}} (\chi) &=& \int^{2\pi}_{0} d\phi \int^{\pi/2}_{0} d\theta \left\{ Q_{1}(\theta,\phi) \sin \theta \right\},\\
    &=& \alpha I_{\rm{in}} \int^{\pi/2}_{0} d\theta \Bigg\{ 
    (-\cos\theta \sin\theta \nonumber + 2\cos\theta \sin^{3}\theta + \cos^{2}\chi \cos^{3}\theta \sin\theta) \left( \int^{2\pi}_{0} d\phi \right) \nonumber\\
    &&\hspace{3.5cm} + 2 \cos\chi \sin\chi \cos^{2}\theta \sin^{2}\theta \left( \int^{2\pi}_{0} \cos\phi d\phi \right) + (\sin^{2}\chi -2) \cos\theta \sin^{3}\theta \left( \int^{2\pi}_{0} \cos^{2} \phi d\phi \right) 
    \Bigg\}, \\
    &=& \alpha \pi I_{\rm{in}} \left\{ -2 \left( \int^{\pi/2}_{0} \cos\theta \sin\theta d\theta \right) + (\sin^{2}\chi +2) \left( \int^{\pi/2}_{0} \cos\theta \sin^{3}\theta d\theta \right) + 2\cos^{2}\chi \left( \int^{\pi/2}_{0} \cos^{3}\theta \sin\theta d\theta \right) \right\},\\
    &=& \alpha \pi I_{\rm{in}} \left( -\frac{1}{2} + \frac{1}{4} \sin^{2}\chi + \frac{1}{2} \cos^{2}\chi \right),\\
    &=& -\frac{\alpha \pi I_{\rm{in}}}{4} \sin^{2} \chi.
\end{eqnarray}

\begin{eqnarray}
    U_{\rm{total}} (\chi) &=& \int^{2\pi}_{0} d\phi \int^{\pi/2}_{0} d\theta \left\{ U_{1}(\theta,\phi) \sin \theta \right\},\\
    &=& -2 \alpha I_{\rm{in}} \int^{\pi/2}_{0} d\theta \left\{ 
    \sin\chi\cos^{2}\theta \sin\theta \left( \int^{2\pi}_{0} \sin\phi d\phi \right)
    - \cos\chi \cos\theta \sin^{3}\theta \left( \int^{2\pi}_{0} \cos\phi \sin\phi d\phi \right) 
    \right\}, \\
    &=& 0.
\end{eqnarray}

Since we assume that all the photons emitted from the unit area are scattered once including the photons that are scattered to the same direction as that before the scattering, the total number of the scattered photons should be the same with that of the emitted photons from the photosphere per unit time ($f_{\rm{in}}$):
\begin{eqnarray}
    f_{\rm{in}} &=& \int^{2\pi}_{0} d\phi \int^{\pi/2}_{0} d\chi \left\{ I_{\rm{total}}(\chi) \sin \chi \right\},\\
    &=& (2\pi I_{\rm{in}}) \alpha \pi \int^{\pi/2}_{0} d\chi \left\{ \frac{3}{2} \sin\chi - \frac{1}{4} \sin^{3}\chi \right\},\\
    &\therefore& \;\;\; \alpha = \frac{3}{4\pi}.
\end{eqnarray}

Therefore, we can calculate the Stokes parameters of the radiation from a unit area on the photosphere using the following simple formula:
\begin{equation} \label{eq:final_Stokes_para}
    \begin{cases}
    I_{\rm{total}} (\chi) = I_{\rm{in}} \left( \frac{9}{8} - \frac{3}{16} \sin^{2} \chi \right),\\
    Q_{\rm{total}} (\chi) = -\frac{3}{16} I_{\rm{in}} \sin^{2} \chi,\\
    U_{\rm{total}} (\chi) = 0.
    \end{cases}
\end{equation}

\end{document}